\documentclass[%
 aip,pop,
 amsmath,amssymb,
 reprint%
]{revtex4-1}

  \usepackage{aas_macros}     

\usepackage{graphicx}
\usepackage{dcolumn}
\usepackage{bm}

\usepackage[utf8]{inputenc}
\usepackage[T1]{fontenc}
\usepackage{mathptmx}
\usepackage{xcolor}

  \renewcommand{\d} {\mathrm{d}}                 

\newcommand{\dash}{\text{--}}

\newcommand{\tA}{\tau_\text{A}}
\newcommand{\tNL}{\tau_\text{nl}}

\begin{document}


\title[Anisotropy and Timescales in 3D MHD Turbulence]{A Detailed Examination of Anisotropy and 
Timescales in Three-dimensional Incompressible 
Magnetohydrodynamic Turbulence}

\author{Rohit Chhiber}  \email{rohitc@udel.edu}
  \affiliation{Department of Physics and Astronomy, Bartol Research Institute, University of Delaware, Newark, DE 19716, USA}
  \affiliation{Heliophysics Science Division, NASA Goddard Space Flight Center, Greenbelt, MD 20771, USA}
  
\author{William H. Matthaeus}%
  \affiliation{Department of Physics and Astronomy, Bartol Research Institute, University of Delaware, Newark, DE 19716, USA}

\author{Sean Oughton}
\affiliation{Department of Mathematics and Statistics, University of Waikato, Hamilton 3240, New Zealand}

\author{Tulasi N. Parashar}
\affiliation{Department of Physics and Astronomy, Bartol Research Institute, University of Delaware, Newark, DE 19716, USA}
\affiliation{School of Chemical and Physical Sciences, Victoria University of Wellington, Kelburn, Wellington 6012, NZ}
  
\date{\today}

\begin{abstract}
When magnetohydrodynamic turbulence evolves in the presence of a large-scale mean magnetic field, an anisotropy develops relative to that preferred direction. The well-known tendency is to develop stronger gradients perpendicular to the magnetic field, relative to the direction along the field. This anisotropy of the spectrum is deeply connected with anisotropy of estimated timescales for dynamical processes, and requires reconsideration of basic issues such as scale locality and spectral transfer. Here analysis of high-resolution three-dimensional simulations of unforced magnetohydrodynamic turbulence permits quantitative assessment of the behavior of theoretically relevant timescales in Fourier wavevector space. We discuss the distribution of nonlinear times, Alfv\'en times, and estimated spectral transfer rates. Attention is called to the potential significance of special regions of the spectrum, such as the two-dimensional limit and the ``critical balance'' region. A formulation of estimated spectral transfer in  terms of a suppression factor supports a conclusion that the quasi two-dimensional fluctuations (characterized by strong nonlinearities) are not a singular limit, but may be in general expected to make important contributions.
\end{abstract}

\maketitle
\section{\label{sec:intro}Introduction}
In standard Kolmogorov theory, turbulence is assumed to be isotropic so that all relevant correlation functions and 
their spectral decompositions are invariant under proper rotations. \citep{Batchelor1953book,Monin1971book} Magnetohydrodynamics (MHD), and magnetized plasma turbulence,
are different in that their dynamics can become anisotropic when a large-scale magnetic field is present. \citep{iroshnikov1964SovAst,kraichnan1965PoF,robinson1971PhFl,
shebalin1983JPP} At this point the assignment of characteristic timescales required for turbulence closures (e.g., Refs \citenum{edwards1964JFM,yokoi2008JTurb}), or even the heuristic identification of timescales in turbulence phenomenologies \citep{kraichnan1965PoFldia,goldreich1995ApJ,zhou2004RMP} becomes more difficult, and even somewhat ambiguous
from a theoretical perspective. How are timescales distributed in wavevector space when the energy is anisotropically distributed? What does scale locality imply when turbulence is anisotropic? What are the relevant relationships among the  different theoretically constructed timescales? Here we will discuss several issues that come into play regarding these timescales, favoring physical relevance over complete mathematical rigor.

We begin with the idea of nonlinear timescale $\tau_\text{nl}$, which is traditionally defined in terms of the scalar magnitude of wavevector, in view of isotropy along with Kolmogorov's assumption of scale locality. The structure of the usual estimate arguably remains valid, even for anisotropic spectra.\citep{zhou2004RMP} Within the inertial range, the time dependence is dominated typically by the random sweeping of inertial-range fluctuations by large eddies, \citep{kraichnan1964pof,chen1989PoF,nelkin1990PoF} both in hydrodynamics \citep{sanada1992pof} and in MHD. \citep{servidio2011EPL} However, this does not directly influence spectral transfer or the nonlinear time, since it corresponds, essentially, to a local spatial translation. \citep{kraichnan1964pof,zhou2004RMP} Another timescale of relevance in MHD, and one that has been the subject of intense discussion, is the Alfv\'en timescale, associated with small-amplitude propagating waves, or under certain specific conditions, large-amplitude fluctuations.\citep{barnes1979inbook} More generally than in the propagating wave scenario, wave-like couplings 
associated with the Alfv\'en timescale act to suppress nonlinear behavior. This occurs due to the potentially rapid changes in the phases of Fourier components that are induced by Alfv\'enic couplings.\citep{dmitruk2009PoP,
bandyopadhyay2018PRE} This effect, which occurs in an anisotropic fashion, reduces the third-order correlations necessary for spectral transfer.
The competition that exists between Alfv\'en and nonlinear effects, as well as the balance of local and nonlocal effects, are issues that have pervaded discussions of MHD and plasma turbulence for decades.  

Here we examine these issues directly by employing MHD simulation. We evaluate the basic timescales throughout Fourier space in a simulation with anisotropy induced by a large-scale mean magnetic field. For clarity of presentation, just one run, employing the incompressible model, and with a fixed uniform externally-supported magnetic field, is the basis for most of what is reported here. However, the results, while not ``universal'' in any meaningful sense,  are expected to remain relevant for useful and familiar regions of parameter space, such as low cross helicity, small magnetic helicity, near-unit Alfv\'en ratio, unit magnetic Prandtl number, etc (see Ref. \citenum{wan2012JFM697} for details). One additional simulation, with a variation in initial data, is discussed in Appendix~\ref{sec:app_iso}. 
We also employ standard methods to estimate quantities related to turbulence activity, such as contributions to spectral transfer in different regions of wavevector space. This strategy will enable an evaluation not only of the instantaneous kinematic state of the energy spectrum \citep[a focus of many prior studies; e.g.,][]{bruno2013LRSP,shalchi2009} but also the greater or lesser concentrations of dynamic turbulence activity, a subject less well-described previously. The results obtained with this strategy will be employed to inform discussion of several approaches to explaining anisotropic spectral distributions of energy, such as multi-component models, \citep{montgomery1981PoF,matthaeus1990JGR,zank1993nearly,
bieber1996dominant} Reduced-MHD (RMHD) models, \citep{strauss1976PoF,montgomery1982PhysScrip,zank1992JPP,schekochihin2009ApJS182}
critical balance,  \citep{goldreich1995ApJ,cho2000ApJ,maron2001ApJ,lithwick2007ApJ,
mallet2015MNRAS} and diffusion models, \citep{matthaeus2009PRE} all of which require consideration of elements of the classic isotropic model for context.\citep{Batchelor1953book,
Monin1971book,frisch1995book}

The outline of the paper is as follows. We briefly describe the simulation used in Section \ref{sec:sims} and present basic results on spectra and timescales in Section \ref{sec:result_basic}. The wavevector distribution of energy and  the so-called \textit{nonlinearity parameter} are examined in Section \ref{sec:result_nltime}. In Section \ref{sec:res_NLrate} we employ these quantities to estimate spectral transfer rates for anisotropic MHD, and examine their spectral distribution as well as their distribution across the nonlinearity parameter. We conclude with discussion in Section \ref{sec:conclude}.
Appendix~\ref{sec:app} provides background to motivate our formulation of the anisotropic spectral transfer rates while Appendix~\ref{sec:app_iso} briefly describes results from a run initialized with fully isotropic  fluctuations.

\section{\label{sec:sims}Simulation Details}
We employ a pseudospectral three-dimensional (3D) incompressible MHD code with \(1024^3\) resolution in a periodic box of dimensionless side $2\pi$  for this study, where the unit length, $L_0$, is arbitrary. \citep{ghosh1993CoPhComm,canuto2012spectral} Our results are primarily obtained from an undriven, freely decaying run whose initial fluctuations are toroidally polarized (i.e., in the same sense as linearized Alfv\'en waves), with some, less detailed references to a second run for which the initial fluctuations are isotropically polarized (toroidal \emph{and} poloidal modes excited). Here we focus on a single run with 
a mean magnetic field \(\bm{B}_0 = 1\hat{\bm{z}}\) (in Alfv\'en units), and initial total energy and Alfv\'en ratio both equal to unity. This corresponds to an initial value of $\delta b/B_0 = 1$ 
where $\delta b$ 
is the root mean square magnetic fluctuation.

Toroidally polarized Fourier modes are excited initially in a band \(3 \le k \le 7\) in units for which the longest allowed wavelength in the periodic box corresponds to a wavenumber of unity. This initial population has a spectral knee at \(k=3\) and at higher $k$ an abbreviated  \(-5/3\) spectral slope. Here \(k\) is the magnitude of the wavevector \(\bm{k}\). Initial cross helicity and magnetic helicity are both very close to zero. The simulation is run using a second-order Runge-Kutta scheme for 8000 timesteps with a step size of \(2.5 \times 10^{-4}\). Therefore the entire run proceeds for about two large-scale nominal nonlinear times. The kinematic viscosity $\nu$ and resistivity $\mu$ are assigned equal values of \(7.3 \times 10^{-4}\). These represent reciprocal Reynolds number and magnetic Reynolds number, respectively, at scale $L_0$.

We emphasize that the results presented here are based on a single snapshot from this simulation. We have repeated the analysis for a similar run initialized with isotropic fluctuations (toroidal + poloidal polarization) and obtained consistent results, discussed below. Numerous other simulations of a similar nature were carried out and were found to present similar properties with regard to resolution, energy decay, and in general the dynamical features of decaying, unforced MHD.\citep{bandyopadhyay2018prx} We therefore believe that the present results are quite typical. However, we caution the reader that true universality is not expected in MHD turbulence due to the multiplicity of controlling parameters (see Ref. \citenum{wan2012JFM697}), and therefore extrapolation of the present observations to widely varying parameters should be undertaken with caution. Further relevant details are presented in the subsequent sections below.

\section{\label{sec:result_basic} Spectra and Basic Timescales}
\begin{figure}
\centering
\includegraphics[scale=.5]{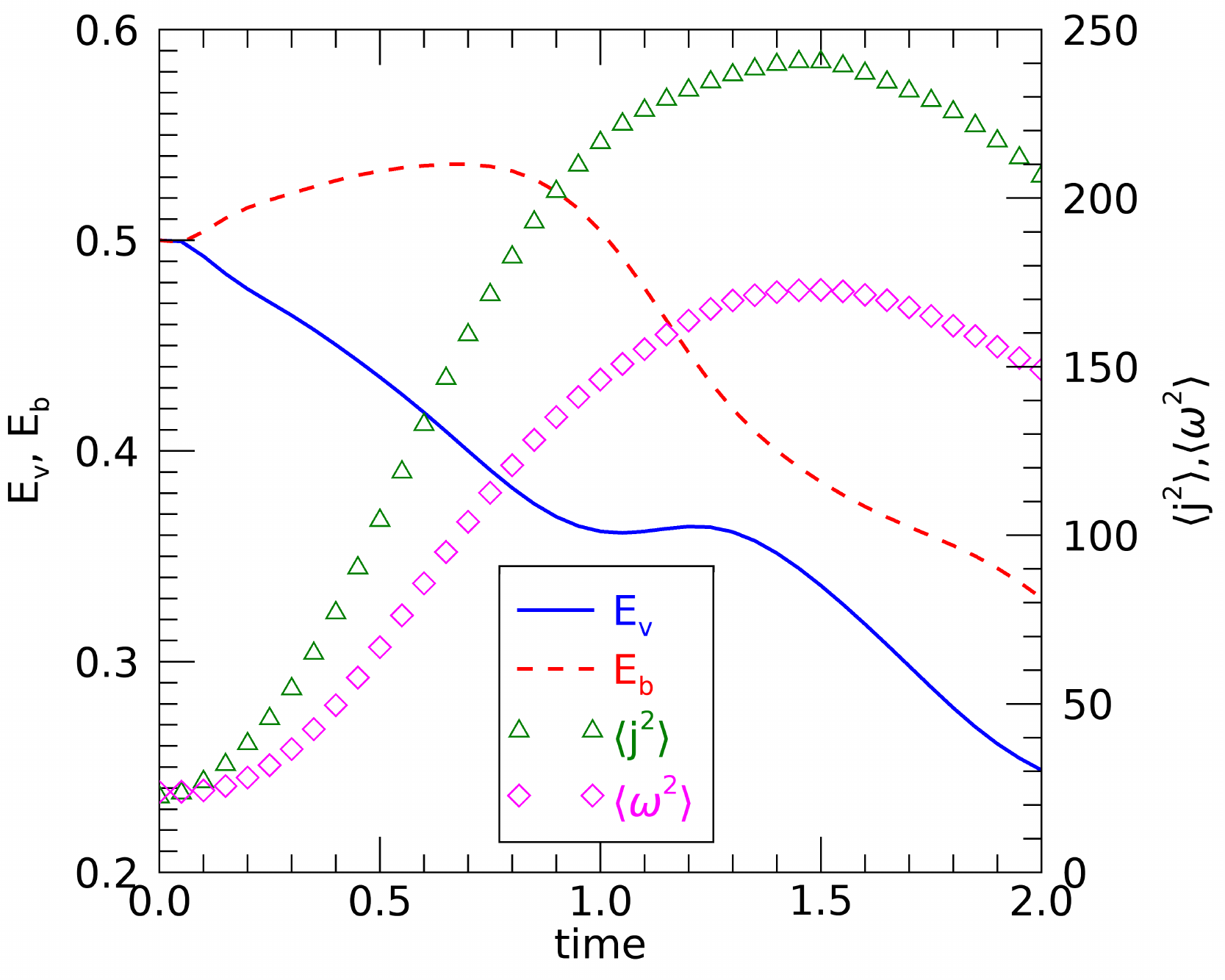}
\caption{Time evolution of kinetic and magnetic energies, mean current density, and enstrophy (\(E_v,~E_b,~\langle j^2\rangle,~\text{and}~\langle \omega^2\rangle\), respectively).}
\label{fig:j}
\end{figure}

\begin{figure}
\centering
\includegraphics[scale=.5]{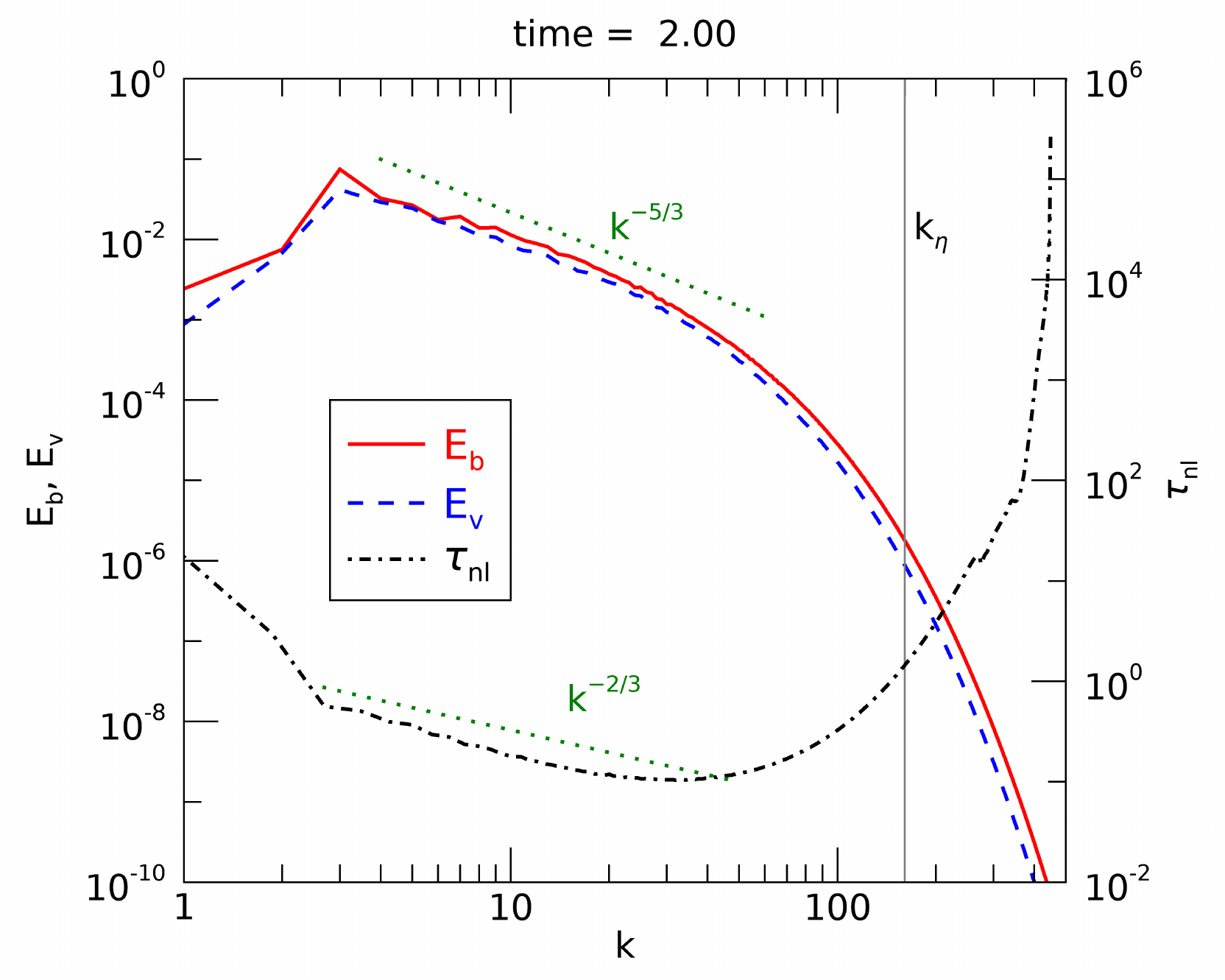}
\caption{Omnidirectional spectra of magnetic (\(E_b\)) and kinetic (\(E_v\)) energies, and nonlinear timescale \(\tNL\), at \(t=2.0\). Here \(\tNL(k) = 1/[k\sqrt{k E_\text{omni}(k)}]\) (see text). The dissipation wavenumber \(k_\eta\) is shown as a vertical line. Reference lines with \(k^{-5/3}\) and \(k^{-2/3}\) slopes are plotted in dotted green.}
\label{fig:spect_t2pt0}
\end{figure}

The context of the following analysis is established by 
examining the time history of key global quantities, 
and wavenumber spectra, which are shown in 
Figures \ref{fig:j} and \ref{fig:spect_t2pt0},
respectively. The evolution of kinetic and magnetic energies (\(E_v\) and \(E_b\), respectively), mean-squared electric current density (magnetic enstrophy) \(\langle j^2\rangle\), and the kinetic enstrophy \(\langle\omega^2\rangle\) are illustrated in Figure \ref{fig:j}.
The enstrophies both peak near about $t=1.5$, 
and accordingly this corresponds to the time of peak dissipation
rate. By $t=2$ we expect that the turbulence is fully developed and that the expected period of von K\'arm\'an similarity decay is established. \citep{karman1938prsl,wan2012JFM697,
bandyopadhyay2018prx} Therefore we carry out the remainder of our analyses at this time,at which the value of $\delta b/B_0$ is about 0.8.
 
 Figure \ref{fig:spect_t2pt0} shows the omnidirectional wavenumber spectra of magnetic energy $E_b$ and kinetic energy $E_v$ at $t=2$. We define the omnidirectional spectrum of total energy as \(E_\text{omni} (k) \equiv E_v(k) + E_b(k)\), so that the total energy per unit mass is \(\int E_\text{omni}(k) \, \d k\). An estimate of the MHD Kolmogorov dissipation wavenumber $k_\eta = \langle j^2 + \omega^2 \rangle^{1/4}/\sqrt{\nu}$ \citep{biskamp2003magnetohydrodynamic} is marked on the plot as a vertical line. An inertial range with an approximate powerlaw slope of $-5/3$ is apparent, extending from about $k=7$ to about $k=30$, falling short of $k_\eta$ by a factor of 2-4 as is typical of fluid models with scalar dissipation coefficients (e.g., Refs.\citenum{pope2000book,ishihara2009AnRFM,
 wan2012ApJ}). The steepening of the inertial-range spectrum becomes significant at about $k=50$, and the spectrum has dropped substantially, by approximately two orders of magnitude, prior to reaching the dissipation wavenumber near $k=160$. We note that this gradual fall-off of the spectrum, and the lack of a clear spectral “break” are typical in well-resolved decaying MHD simulations, \citep{wan2010PoP} in contrast to the rather sharper termination of the spectrum often seen in the solar wind (e.g., Ref. \citenum{leamon1998JGR103}). This is because the MHD fluid model in this paper does not capture the collisionless-plasma effects on nonlinear dynamics that come into play at kinetic scales in the solar wind (eg., Ref. \citenum{Goldstein2015RSPTA}).

%

%
%

To proceed we examine the basic timescales. First we consider the nonlinear timescale, defined as local in wavenumber, following Kolmogorov's assumption of the dominance of local transfer. It is then standard practice to define $\tNL(k) \equiv [k \delta v(k)]^{-1}$, where \(\delta v(k)\) is the characteristic speed at scales \(\ell \approx 1/k\). Using the omnidirectional total-energy spectrum, with \(\delta v(k)^2 \approx k E_\text{omni}(k)\), yields 
\begin{equation}
\tNL(k) = 1/[k\sqrt{k E_\text{omni}(k)}].
\label{eq:tNL}
\end{equation}
This definition maintains the plausible approximation that the nonlinear couplings are mainly local in the magnitude of wavevector.\citep{pouquet1976JFM,domaradzki1990PoF,
zhou2004RMP,
eyink2005PhyD,verma2005PoP,aluie2010PRL}
 
 The dependence on $|\bm{k}|$ of this standard, scale-local nonlinear time is shown in Figure \ref{fig:spect_t2pt0} (see also Refs.\citenum{matthaeus2014apj,papini2019NCimC}). There is no angular dependence with this definition, and nonlinear time so-defined is constant on shells with a given modulus of wave vector. At this modest Reynolds number the inertial range spans only about one decade, as was seen also in the energy spectra. Here the expected inertial range behavior $\tNL(k) \sim k^{-2/3}$ is seen only over a rather narrow approximate range of $k=7$ to $k=35$.  For $k>100$ the scale-local nonlinear time sharply increases due to rapidly decreasing spectral energy density. But in any case the approximation of locality is questionable as the dissipation scale is approached. Similarly, this definition of local nonlinear timescale is not relevant to the lowest few wavenumbers at the energy-containing scales, as the granularity of the excited modes comes into play and the strength of nonlinear couplings depends on the specific energy distribution in that range.

\begin{figure}
\centering
\includegraphics[scale=.6]{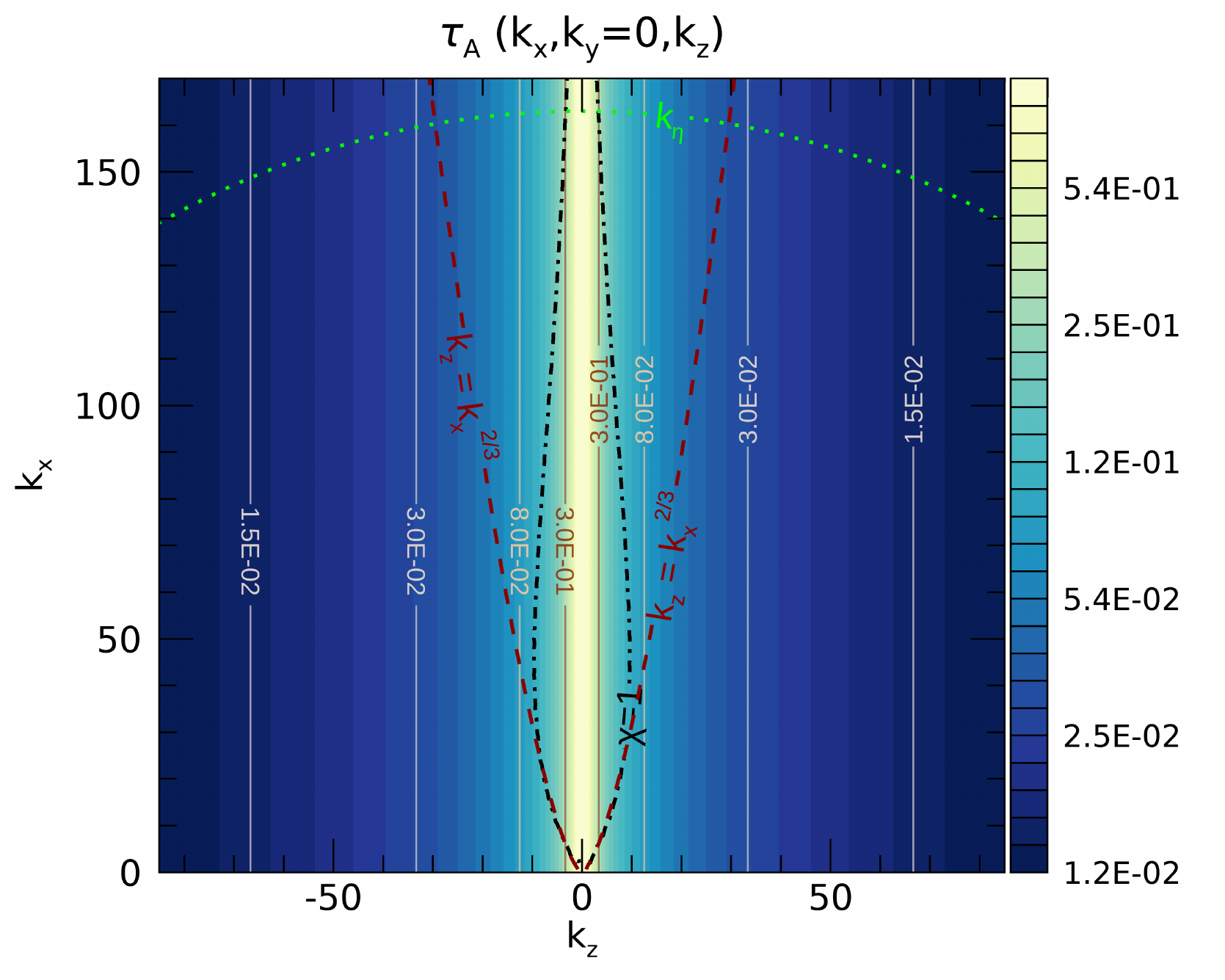}
\caption{Alfv\'en timescale \(\tA(\bm{k}) = 1/|k_z B_0|\) in the \(k_z \dash k_x\) plane. For \(k_z=0\) we set the Alfv\'en time equal to the neighboring value at \(k_z=1\): \(1/B_0\). The Higdon\cite{higdon1984ApJ} curve (\(k_z = k_x^{2/3}\)), the \(\chi= \tA/\tNL = 1\) curve (critical balance\cite{goldreich1995ApJ}), and the dissipation wavenumber \(k_\eta\) are plotted as dashed cyan, dash-dotted black, and dotted green curves, respectively. The latter two contours are constructed from simulation data at \(t=2\).}
\label{fig:t_alf_kperppar}
\end{figure}

The principal antagonist to nonlinear effects in incompressible MHD turbulence with a large-scale magnetic field is the Alfv\'en propagation effect \citep{kraichnan1965PoF} that interferes -- \textit{anisotropically} -- with couplings that drive turbulence. The associated Alfv\'en timescale $\tA(\bm{k}) = 1/|k_z B_0|$ is shown in Figure \ref{fig:t_alf_kperppar}, where the distribution of \(\tA(\bm{k})\) is illustrated over the \(k_z \dash k_x\) plane defined by \(k_y = 0\). Along the $k_z=0$ axis (the 2D axis) the value is formally infinite, but is depicted with the same color intensity as the neighboring $k_z=1$ modes for visualization purposes. Constant Alfv\'en-time contours are parallel to the $k_x$ axis, by definition. The other contours in Figure~\ref{fig:t_alf_kperppar} will be presently discussed below and are included for later referencing. See Ref. \citenum{Ghosh2015PoP1} for similar results based on Fourier-space ratios of linear and nonlinear accelerations, rather than timescales.

\section{\label{sec:result_nltime} Wavevector distribution of energy and nonlinearity parameter}
\begin{figure}[h]
\centering
\includegraphics[scale=.56]{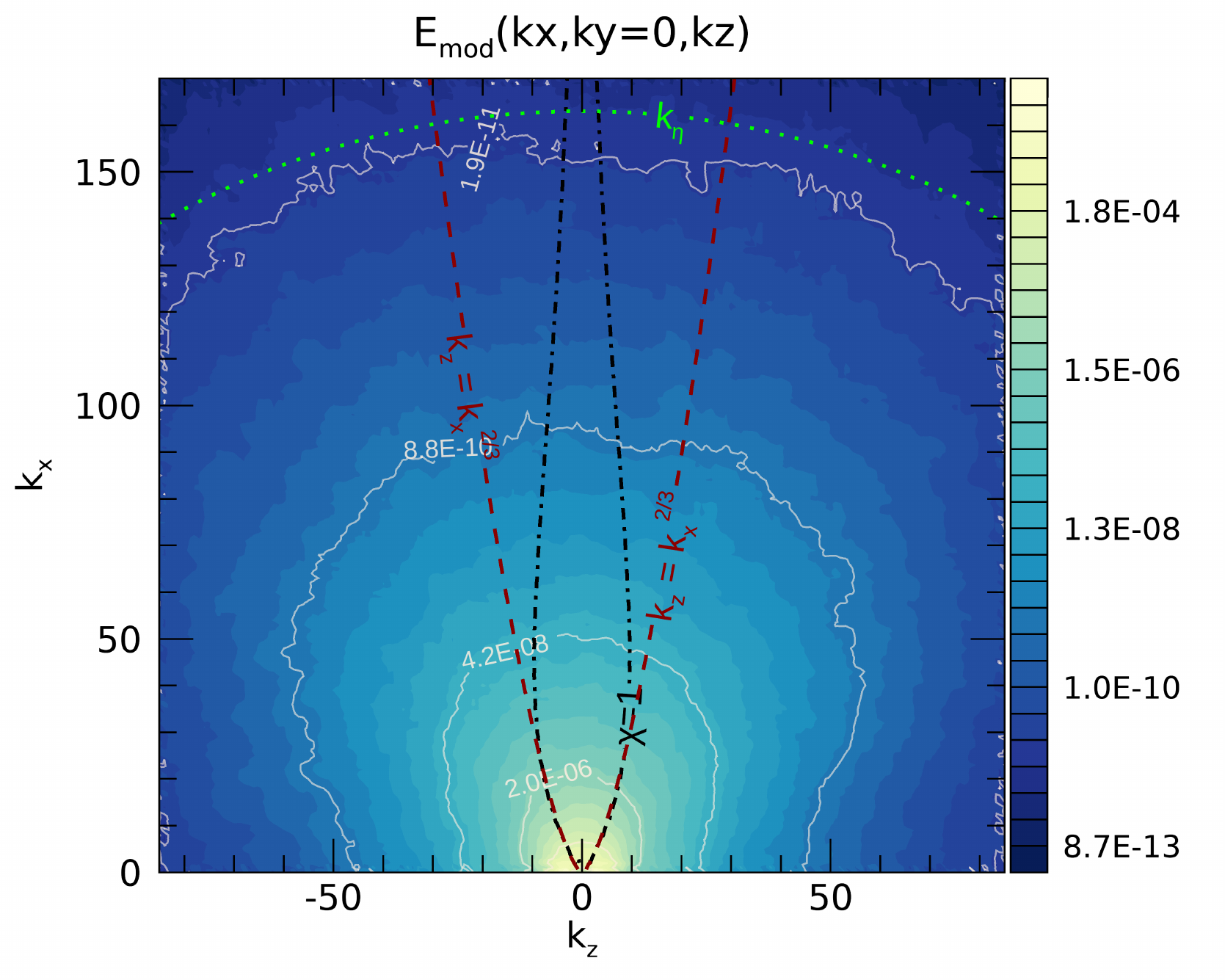}
\includegraphics[scale=.56]{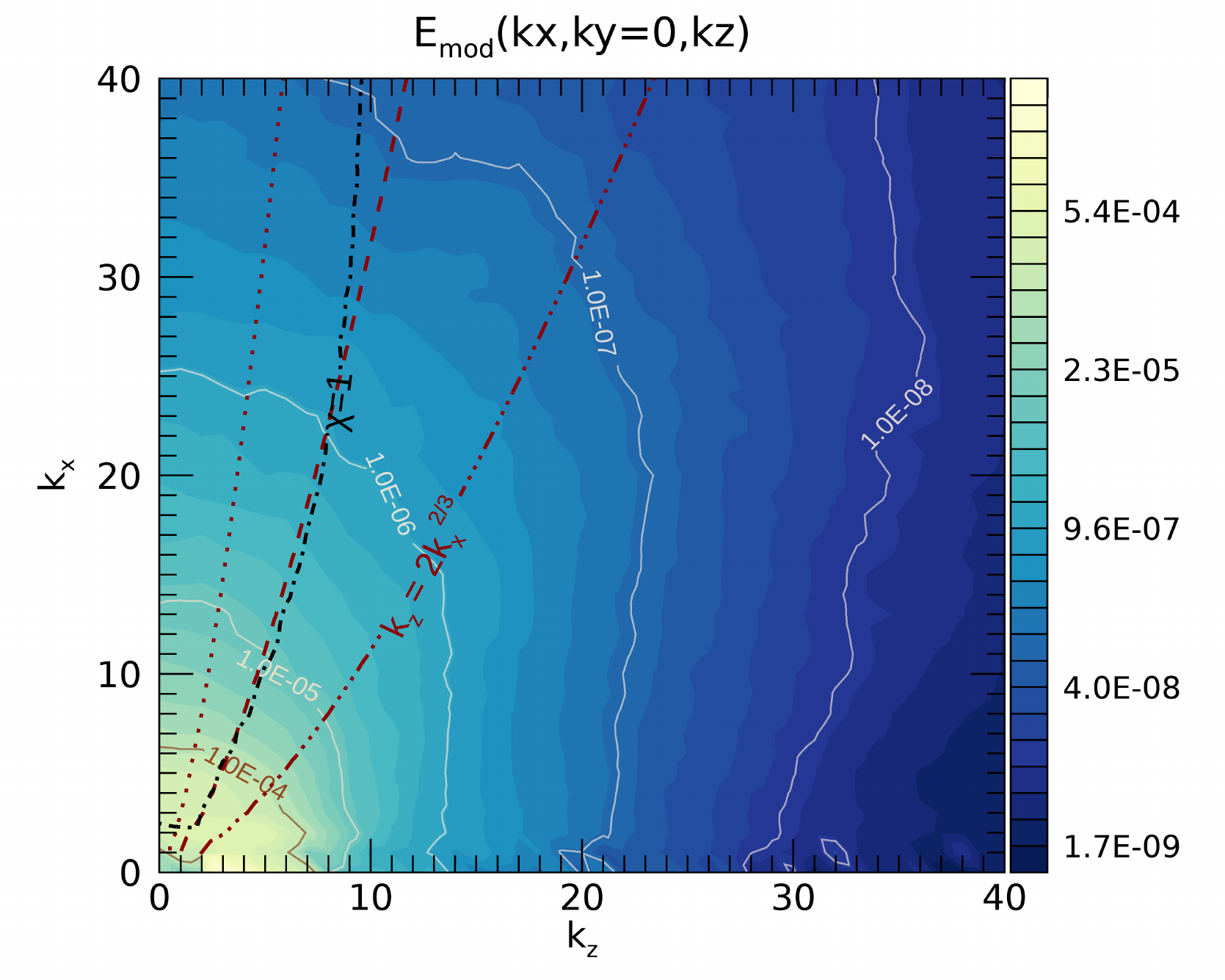}
\caption{\textit{Top:} Contours of modal energy spectrum in the \(k_z\dash k_x\) plane. The Higdon curve, the \(\chi=1\) curve, and the dissipation wavenumber are plotted as dashed dark-red, dash-dotted black, and dotted green curves, respectively. \textit{Bottom:} A blow-up of the region near the origin, covering the inertial range of wavenumbers. The dark-red dotted, dashed, and dash-triple-dotted lines show Higdon curves with proportionality constant \(\alpha\) equal to 0.5, 1, and 2, respectively.}
\label{fig:Emod}
\end{figure}

A more detailed view of the energy spectrum is afforded by 
examining the distribution of energy in parallel and perpendicular components of wavevector. We define the modal energy spectral density as \(E_\text{mod}(\bm{k}) = [|\bm{v}(\bm{k})|^2 + |\bm{b}(\bm{k})|^2]/\Delta k^3\), where \(\bm{v}(\bm{k})\) and \(\bm{b}(\bm{k})\) are the (periodic domain) Fourier coefficients of velocity and magnetic field,  respectively, and \(\Delta k^3\) is the fundamental Cartesian volume element for the simulation (equal to unity in our case). Figure \ref{fig:Emod} shows a map of \(E_\text{mod}(\bm{k})\) in a two-dimensional (2D) cut through the wavevector space, for the simulation at $t=2.0$. The particular selected plane, $E_\text{mod}(k_x,k_y=0,k_z)$, spans one perpendicular direction, $k_x$, and the parallel direction, $k_z$. In the Figure, lines of constant spectral density are shown, the pattern of distribution further emphasized by plotting colors associated with ranges of energy density. The following analyses and figures correspond to \(t=2.0\) as well.

The top panel shows a large range of wavevectors, extending beyond the dissipation scale in the high $k_x$ (perpendicular)
direction, and spanning about ten orders of magnitude in spectral density. The bottom panel shows a close-up of the same distribution, concentrating on the region closer to the origin
with $k_x\le 40$ and $k_z\le 40$, covering the inertial range.

The distribution is highly anisotropic, extending more deeply towards the perpendicular direction, as expected given the well-known tendency for MHD turbulence to produce stronger gradients perpendicular to an externally-supported uniform magnetic field. \citep{robinson1971PhFl,shebalin1983JPP,
oughton1994JFM} One may note, for example, that the contour labeled as $2.9 \times 10^{-10}$  extends beyond $k_x=100$ at $k_z=0$ while the same contour, at $k_x=0$, extends only to about $k_z=55$.

The bottom panel of Figure \ref{fig:Emod} emphasizes the range of wave vectors that would typically be included in the inertial range. The contours are almost vertical at high $k_z>25$ and indicate peaks in the spectral density at $k_z=0$ for all values of $k_x$. Therefore one immediately concludes that the ``quasi-2D'' modes ($k_z \ll k$) are the most energetic, and in some ways may be dominant. We explore this further below. Note that, in general, for MHD one would define quasi-2D modes as those with $k_z \ll k$  and  $ \tNL (k) \ll \tA (k)$.\cite{oughton2005NPG} In a continuum there may be quasi-2D modes with nonzero $k_z \ll k$  (see Figure \ref{fig:kdiagram}), but for our discrete case and our particular simulation parameters, the quasi-2D requirements are satisfied only for the $k_z=0$ modes.

Also plotted in the panels of Figure \ref{fig:Emod} are contours on which the \textit{computed} nonlinearity parameter, 
\begin{equation}
\chi(\bm{k}) = \frac{\tA(\bm{k})} {\tNL(k)},
\label{Chi}
\end{equation}
takes on the values of unity. This parameter provides a measure of the relative strength of local-in-scale nonlinearity and local Alfv\'en wave propagation (see, e.g., Refs. \citenum{montgomery1982PhysScrip,
goldreich1995ApJ,
mallet2015MNRAS}). This idealized ``equal-timescale curve'' is computed assuming inertial-range Kolmogorov scaling, associated with the Higdon\cite{higdon1984ApJ} formula $k_z L_0 \sim (k_xL_0)^{2/3}$, and was later adopted by Goldreich \& Sridhar\cite{goldreich1995ApJ} as a central element of the critical balance phenomenology. The contours of $k_z = \alpha k_x^{2/3}$ (in dimensionless units), associated with \(\tNL = \text{constant}\times\tA\), are also shown for different values of the proportionality constant \(\alpha\) (see caption).
 
A significant feature of the spectra in Figure \ref{fig:Emod} is the \emph{lack} of any apparent preferred regions in wavevector space. In particular, there is no discernible enhancement of energy density near the Higdon/critical-balance curve, or along any curve of that order, such as the $k_z = 2 (k_x)^{2/3}$ curve that is illustrated (see also discussion of Figures 2 and 3 in Ref. \citenum{Ghosh2015PoP1}). We also note that there is no discernible \emph{deficit} of spectral power along either the quasi-2D axis $k_z=0$ or along the ``slab'' axis $k_x=0$. Note that the latter observation is pertinent to all values of wavenumber, not only the initially excited range including $|\bm{k}|<7$. Therefore, for example, the small-scale quasi-2D modes are populated by the cascade on the same timescales as those over which the rest of the anisotropic spectral distribution is formed.

\begin{figure}[h]
\centering
\includegraphics[scale=.6]{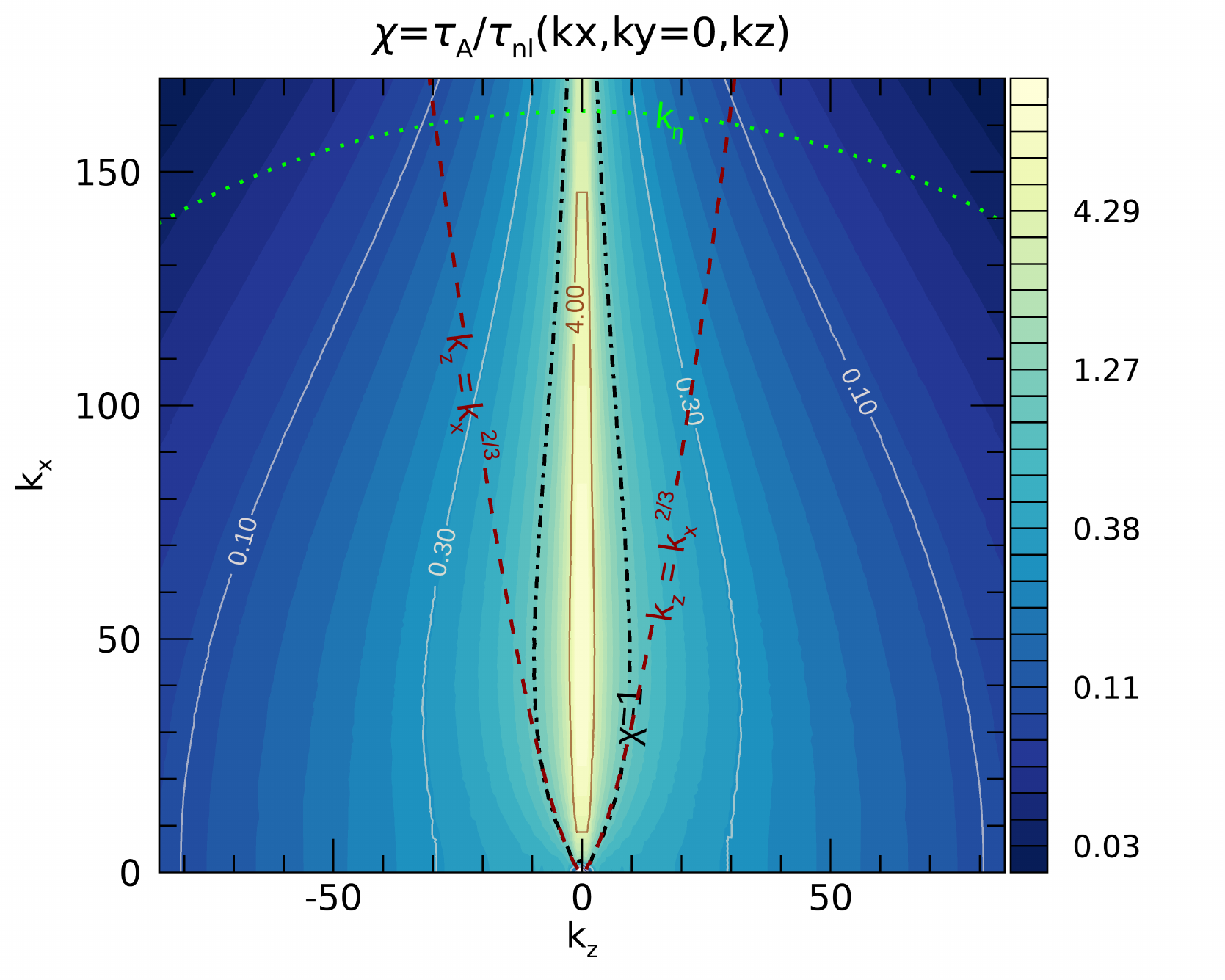}
\includegraphics[scale=.6]{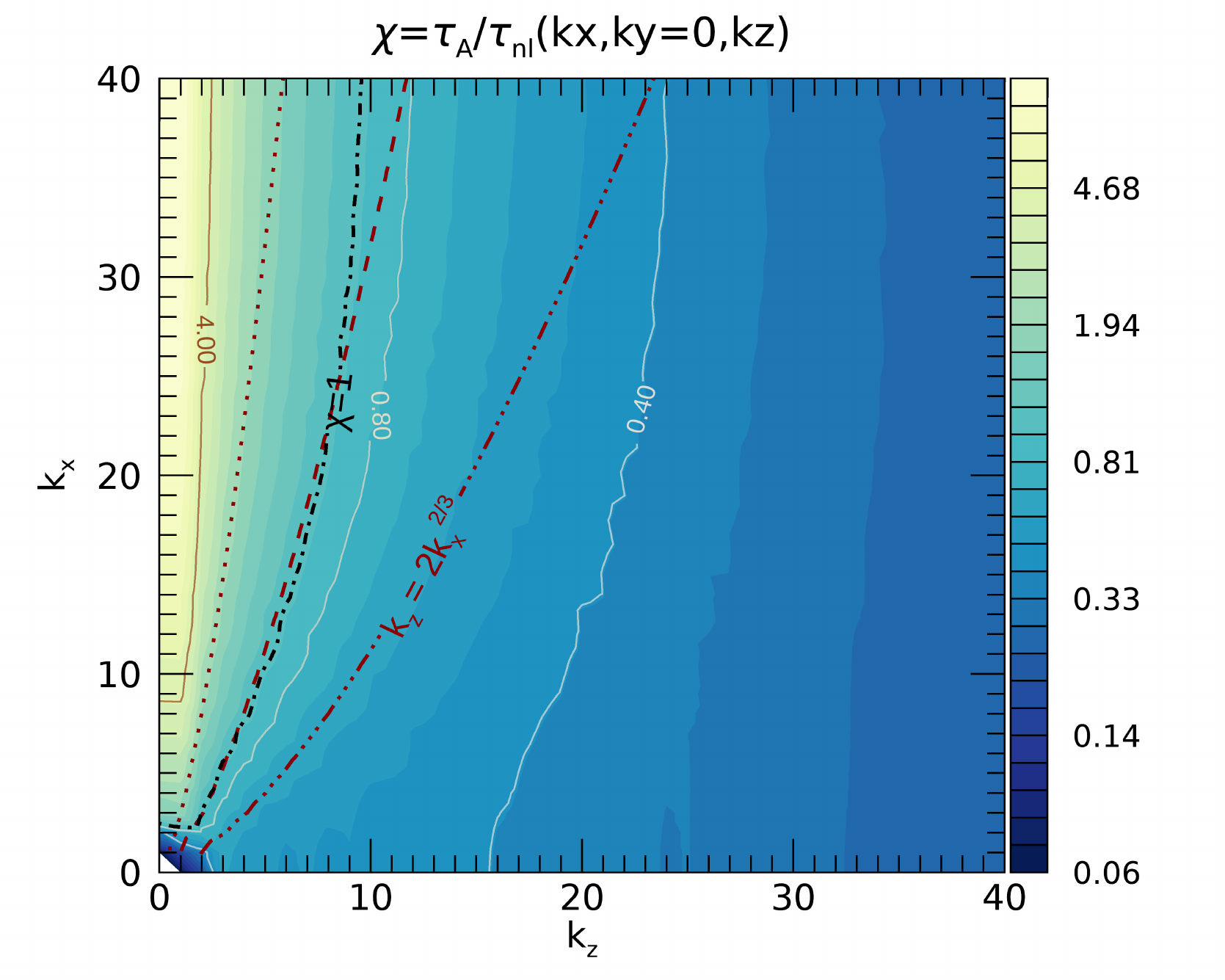}
\caption{\textit{Top:} Nonlinearity parameter \(\chi = \tA/\tNL\). \textit{Bottom:} A blow-up of the region near the origin. In both panels, the dissipation wavenumber and the \(\chi=1\) and Higdon curves are plotted as described in Figure \ref{fig:t_alf_kperppar}.}
\label{fig:chi}
\end{figure}

Figure \ref{fig:chi} provides complementary information by illustrating the distribution of the nonlinearity parameter in the same 2D cut through $k$-space. Here the nonlinearity parameter is computed from the actual spectral distribution, and not from an assumed Kolmogorov powerlaw. As in the previous figure, this accounts for why the computed $\chi = 1$ line does
not coincide (outside the inertial range) with the idealized Higdon curve, which does adopt the assumption of a canonical $-5/3$ powerlaw spectrum in the perpendicular wavenumber. 

One may observe in Figure \ref{fig:chi} that there is a concentration of values of $\chi$ greater than unity near the $k_z=0$ axis, i.e., at or near the quasi-2D region of $k$-space. Here too the formally infinite values of $\chi$ on the 2D axis have been replaced with values found at the juxtaposed modes with \(k_z=1\). We recall then that the energy density is substantial in these modes, and actually somewhat larger than modes found at higher $k_z$ closer to the $\chi=1$ curve. This provides a clear suggestion of the difficulty in asserting that $\chi=1$ is a kind of upper attainable limit of the nonlinear strength. However this is not a quantitative statement, as it falls short of an assessment of the relative importance of regions of the spectrum with different $\chi$ values. Analysis of that type is provided below.  
 
\section{\label{sec:res_NLrate} Nonlinear transfer-rate estimates}

In the previous section we examined the anisotropic distribution of energy due to an externally supported mean magnetic field, and we computed the basic physical timescales at the instant at which the analysis is performed. To proceed to an understanding of the dynamics, one must go further. Several options exist.

One approach is to compute, based on the known state of
the system, the scale-to-scale transfer rates using a scale-filtering approach.\citep{germano1992JFM} Studies of this type have been done for isotropic MHD; \citep{alexakis2005PRE,verma2005PoP,
aluie2010PRL} the extension of such studies to the anisotropic case has been less frequently studied. However, for the anisotropic case, there are some models based on physical assumptions consistent with those adopted here, \citep{alexakis2007PRE,matthaeus2009PRE} while examination of transfer across surfaces such as planes and cylinders also led to conclusions \citep{alexakis2007PRE} broadly consistent with the ideas adopted here. 

We note that while computing scale-to-scale transfer directly is extremely valuable for addressing questions such as the validity of Kolmgorogov locality, it does not immediately provide insights into the role of the various physical timescales in producing the computed effects. In particular, if we wish to develop or test phenomenological theories \citep{montgomery1981PoF,matthaeus1989PoF,goldreich1995ApJ,
biskamp2003magnetohydrodynamic} or closures, \citep{pouquet1978JFM,grappin1982AA,goldreich1995ApJ,
yokoi2008JTurb,yoshizawa2013hydrodynamic} it will be necessary to understand how the fundamental timescales enter the dynamics.

Similar challenges exist for interpretation of the 3D form of the Kolmogorov--Yaglom--Politano--Pouquet laws. \citep{politano1998GRL,osman2011PRL_third,
wan2012JFM697,verdini2015ApJ} This relationship provides a direct connection between third-order correlations and total energy-transfer across an arbitrary closed surface in scale space.\citep{wan2009PoP_shear} However third-order laws are notorious for their slow convergence, and in any case require additional analysis to connect their empirically obtained values to physical parameters and timescales. This is particularly germane here because the applied magnetic field does not change the formal structure of the third-order laws, but rather appears in the hierarchy on the same footing as the fourth-order moments. \citep{wan2012JFM697} 
Consideration of higher-order moments would complicate the present study in which the goal is to quantify the roles of the nonlinear and wave timescales in spectral transfer. 

In view of these issues, in what follows we pursue a different and simpler approach, emphasizing potential insights into the role of the fundamental timescales in inducing spectral transfer of energy. We employ existing frameworks of turbulence phenomenologies, plausibly generalized to the anisotropic case, to probe relative strength of nonlinear and linear effects, and to understand the role of the associated timescales in controlling spectral transfer. 

\subsection{Elementary estimates}
The simplest estimate of energy transfer ignores both the influence of Alfv\'enic propagation and the anisotropic distribution of energy in wavevector. In effect one averages the nonlinear effects and the energy distribution over angle, leaving a dependence on wavenumber alone. Such a direction-averaged estimate of (local) energy spectral-transfer rate may be written as $\delta v^2(k)/\tNL(k) = kE_\text{omni}(k)/\tNL(k)$. In steady state, this rate, in units of energy per unit mass per unit time, will be balanced by the global dissipation rate $\epsilon =\nu\langle j^2 + \omega^2\rangle$. This approach is fully equivalent to the classic Kolmogorov theory, since all information about anisotropy is lost through averaging. Such an approach has been successfully applied to solar wind observations to provide useful estimates of rates of cascade and dissipation. \citep{verma1995JGR,leamon1998JGR103,
vasquez2007JGR} It is not necessary to plot this spectral transfer rate in regions of wavevector space as has been done above in Figure \ref{fig:chi}, since the pattern would consist simply of concentric circular regions. 

\begin{figure}[h]
\centering
\includegraphics[scale=.6]{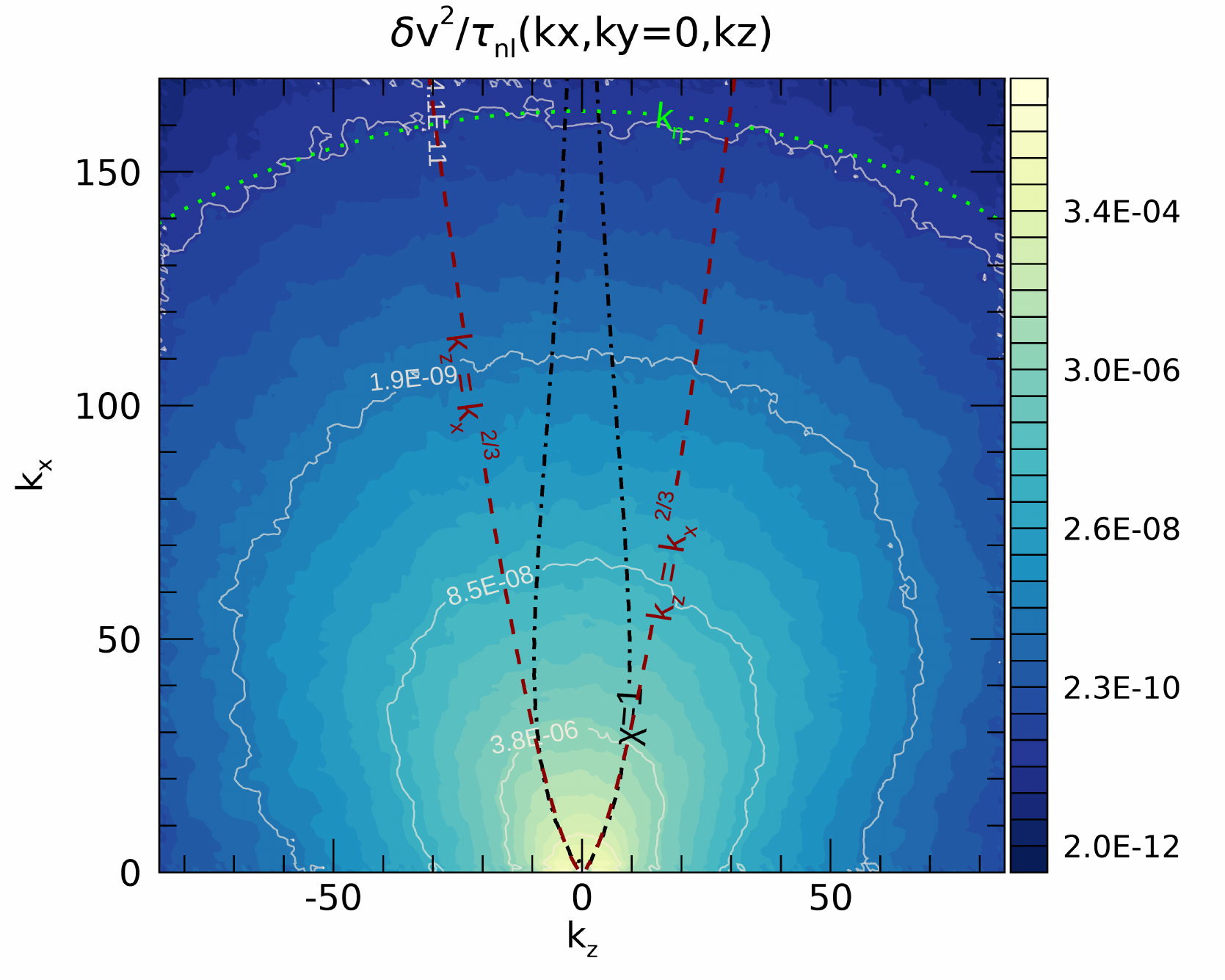}
\includegraphics[scale=.6]{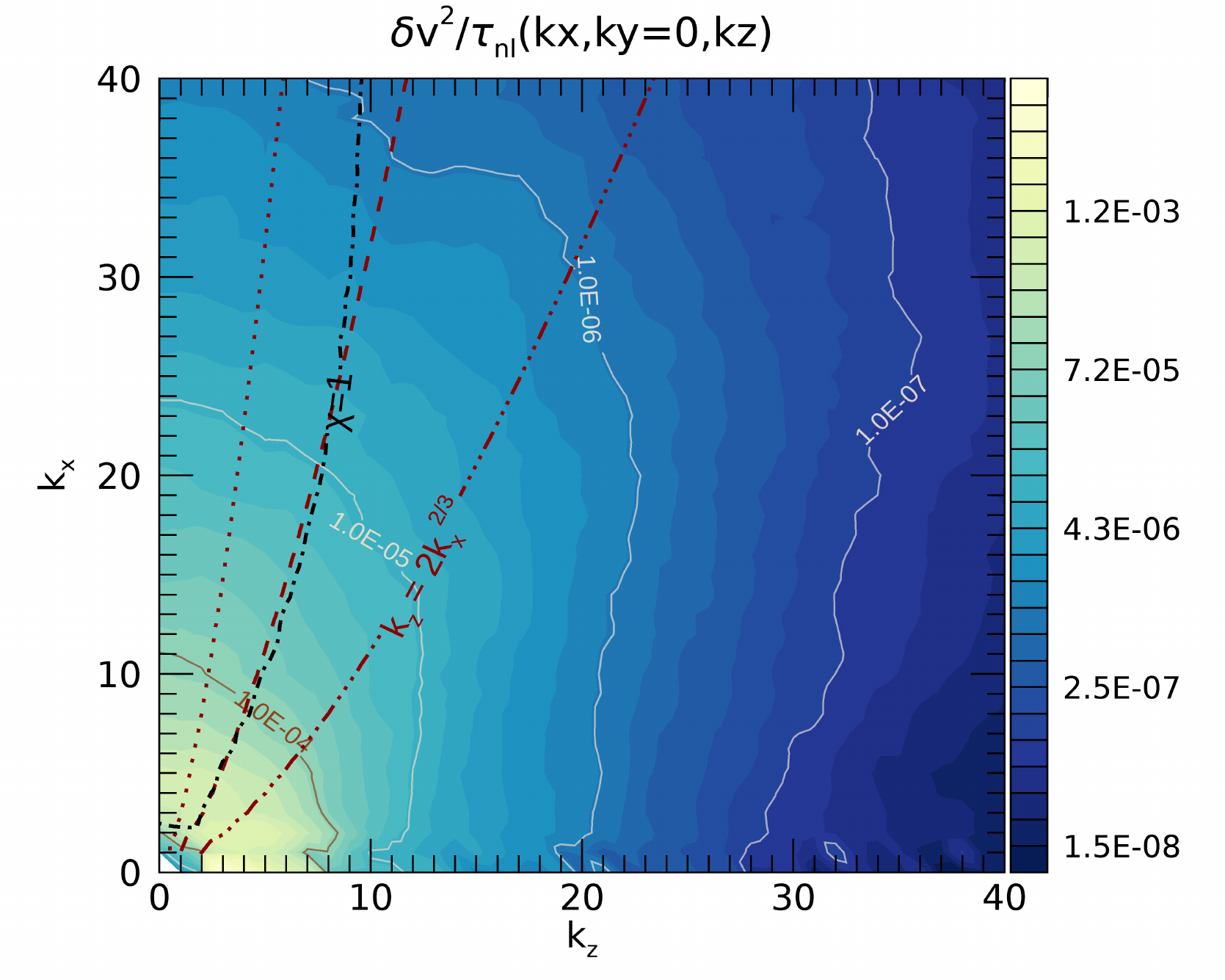}
\caption{\textit{Top}: Estimate of local energy-cascade rate \(\delta v^2(\bm{k})/\tNL(k)\). \textit{Bottom}: A blow-up of the region near the origin. In both panels, the dissipation wavenumber and the \(\chi=1\) and Higdon curves are plotted as described in Figure \ref{fig:Emod}.}
\label{fig:Emod_tnl}
\end{figure}

A modified estimate, also incomplete, is one that includes anisotropy of the energy distribution, but no explicit account of the Alfv\'enic timescale. To take this step, we maintain Kolmogorov's scale-locality in computation of the nonlinear timescale and quantify its effect on the \emph{anisotropic} spectral distribution of energy. This estimate of a ``local pseudo-cascade rate'' may be expressed as 
\begin{equation}
  \frac {\delta v^2(\bm{k})} {\tNL(k)} 
    = 
  \frac {\Delta k^3 E_\text{mod}(\bm{k}) } {\tNL(k)},
\label{estimate1}
\end{equation}
and is illustrated in Figure \ref{fig:Emod_tnl}. This is, in effect, an estimate of spectral transfer acting at a given wavevector if the mean magnetic field is suddenly extinguished while the spectrum remains anisotropic.  In the following we denote the turbulence strength at wavevector \(\bm{k}\) as \(\delta v(\bm{k})\), estimated using Equation \eqref{estimate1}.

Several properties of the quantity plotted in Figure \ref{fig:Emod_tnl} are apparent. The first is that it is anisotropic  in the plane, but also that its anisotropy is identical to that of the modal energy spectrum itself, which was shown in Figure \ref{fig:Emod}. This is due to the fact that the  Kolmogorov nonlinear timescale, as defined in Equation \eqref{eq:tNL}, is isotropic. For this reason, the quantity depicted in Figure \ref{fig:Emod_tnl}, even if it has the dimensions of energy cascade rate (per unit mass), is not useful as a measure of the anisotropy of the cascade. Such a measure must include influence of the Alfv\'en propagation effect (see Figure \ref{fig:t_alf_kperppar}), which is known to suppress spectral transfer in the parallel direction in \(k\)-space.\citep{montgomery1981PoF,shebalin1983JPP}

\subsection{Triple Correlations and Introduction of the Alfv\'en time}

An improved estimate of local cascade rates may be constructed based on timescales. We begin with the formulation of spectral transfer rate in terms of nonlinear time and the lifetime of triple correlations as formulated by Kraichnan\cite{kraichnan1965PoF} and subsequently extended into a ``golden rule''. \citep{matthaeus1989PoF,frisch1995book,
zhou2004RMP} Note that the triple correlations are so named since they involve triple products of Elsasser-field components, and are responsible for inducing turbulent energy transfer across the inertial range. The essential statement, formulated for the isotropic case, is that the spectral transfer time may be defined by the relation \(\tau_\text{sp}(k) \tau_3(k) = \tNL^2(k)\), which is equivalent to Kraichnan's observation that in steady state the transfer rate
\begin{equation}
\epsilon(k) = \tau_3(k) \frac{[\delta v(k)]^2} {\tNL(k)^2}
\label{eq:epsilonk}
\end{equation}
must be independent of scale. In this expression, the direct proportionality on the lifetime of triple correlations $\tau_3$ is a physical requirement, one that may be amply motivated by examining the structure of moment hierarchies appearing in closures such as the Eddy Damped Quasi-Normal Markovian Approximation.\citep{orszag1970JFM,zhou2004RMP} Furthermore, the only other timescale available on the right side of Equation \eqref{eq:epsilonk} is the nonlinear time \(\tNL\). It follows that the effective spectral transfer rate is 
\begin{equation}
\tau_\text{sp}^{-1} = \tau_3/\tNL^2.
\label{tspec}
\end{equation}
At this point we introduce anisotropic effects. While the nonlinear time (assuming scale locality) is a function only of the magnitude $|\bm{k}|$, there is a possibility to introduce on physical grounds a directional dependence in $\tau_3$. This introduces a wavevector directional dependence in the spectral transfer rate $1/\tau_\text{sp}$.

\begin{figure}[h]
\centering
\includegraphics[scale=.56]{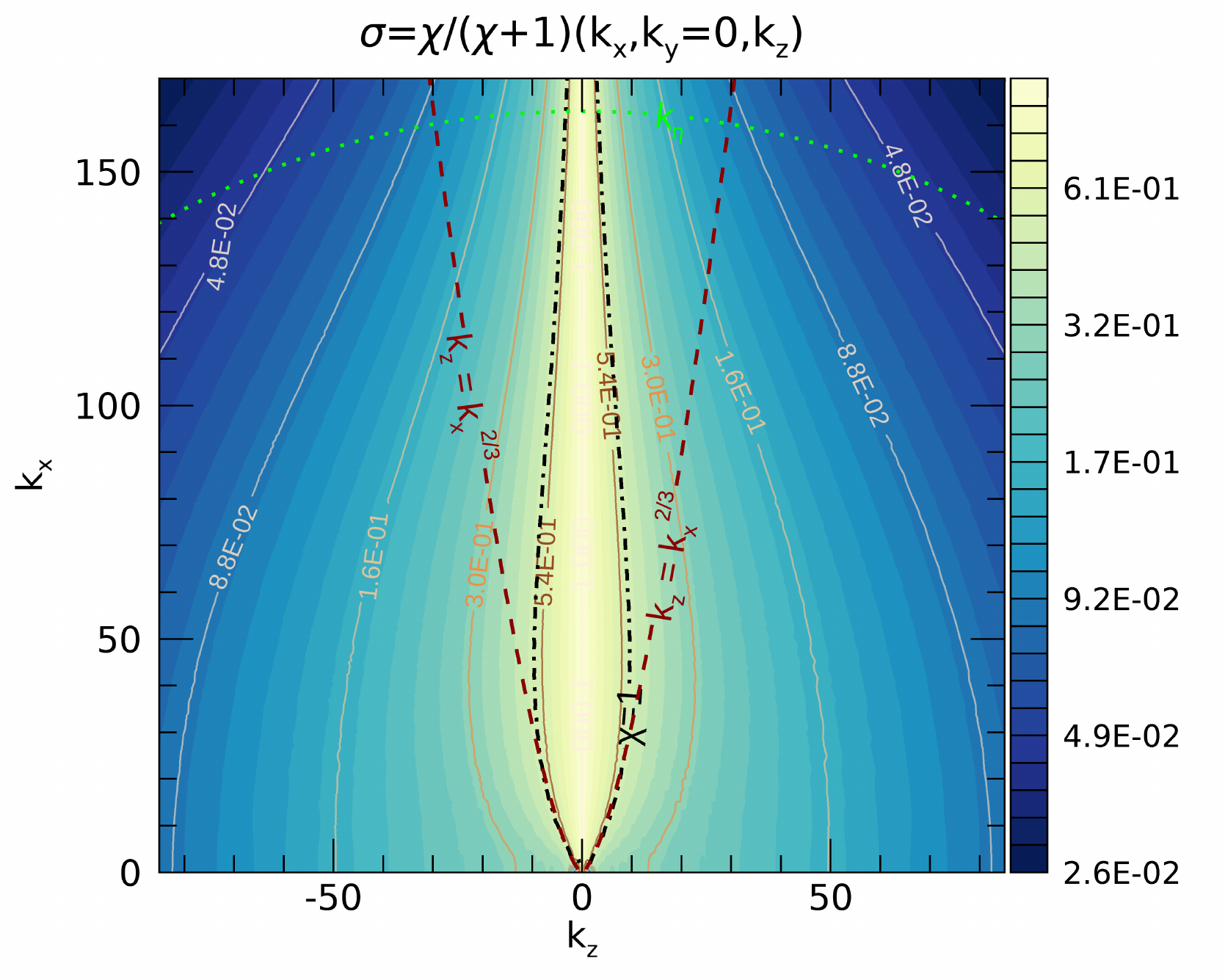}
\includegraphics[scale=.56]{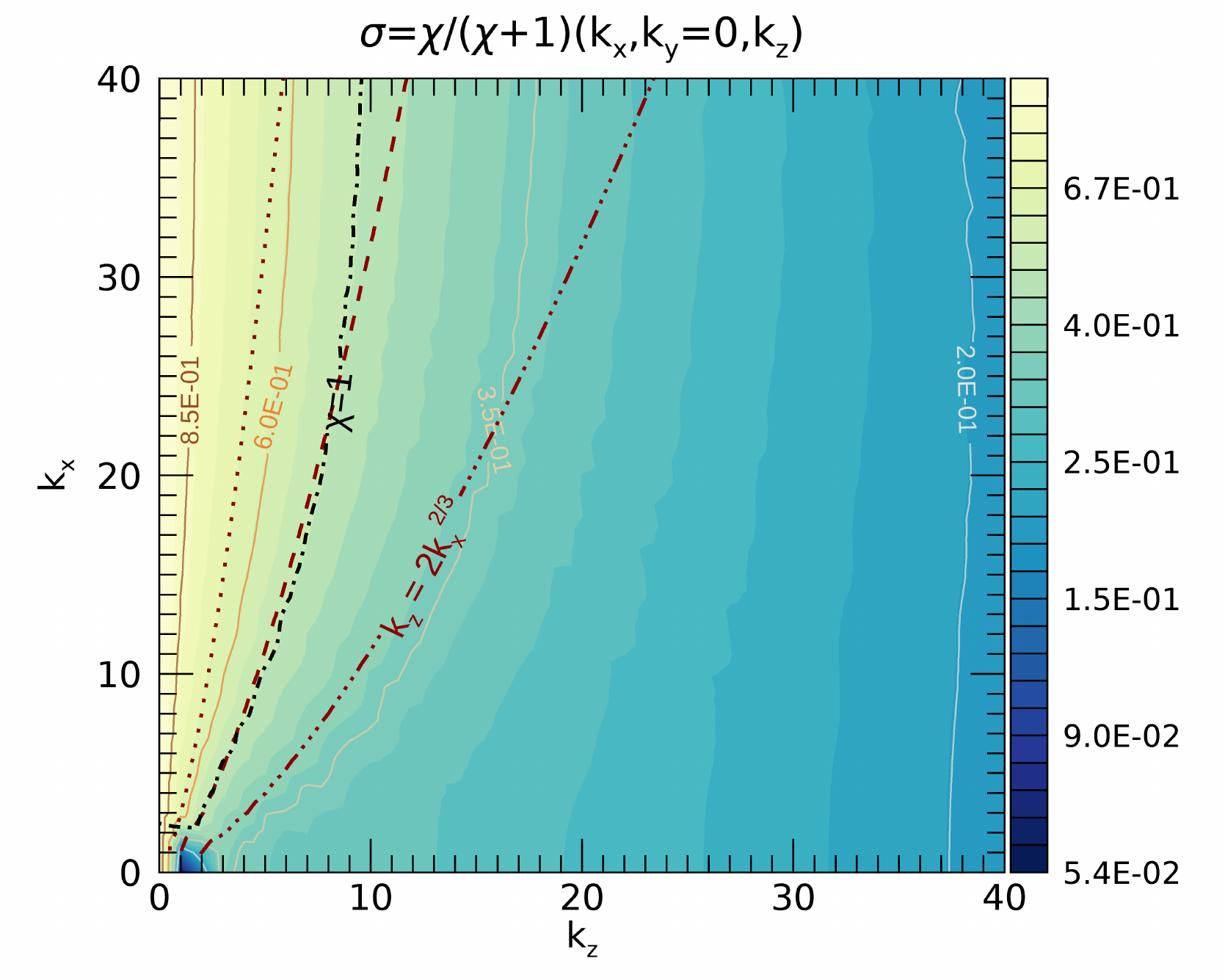}
\caption{\textit{Top}: Suppression factor for anisotropic spectral transfer \(\sigma = \chi/(\chi+1)\). At \(k_z = 0\) we have \(\chi \to \infty\), and therefore we set \(\sigma\) equal to unity there. \textit{Bottom}: A blow-up of the region near the origin. In both panels, the dissipation wavenumber and the \(\chi=1\) and Higdon curves are plotted as described in Figure \ref{fig:Emod}.}
\label{fig:sigma}
\end{figure}

To proceed, we approximate the lifetime of the triple correlations. \citep{kraichnan1965PoF} A standard approach is to assume that the total rate of decay of the triple correlations is the sum of contributions from individual rates. \citep{pouquet1976JFM,matthaeus1989PoF,
zhou2004RMP} Here, the available rates are those derived from nonlinear effects and the Alfv\'en propagation effect. Specifically, allowing for the directionality of Alfv\'en propagation, we may write
\begin{equation}
\frac{1}{\tau_3}=\frac{1}{\tA}+\frac{1}{\tNL},
\end{equation}
which yields 
\begin{equation}
\tau_3 (\bm{k}) = \frac{\tA(\bm{k}) \tNL(k)}{\tA(\bm{k}) + \tNL(k)},
\label{eq:t3}
\end{equation}
where as before, \(\tA(\bm{k}) = (\bm{k} \cdot \bm{V}_A)^{-1} = (k_z B_0)^{-1}\) for Alfv\'en speed \(\bm{V}_{A}\). This leads to a \emph{locally} defined anisotropic spectral-transfer rate (see Appendix~\ref{sec:app}): 
\begin{equation}
\frac{1}{\tau_\text{sp}(\bm{k})}
=
\frac{\tau_3(\bm{k})}{ \tNL^2(k)}
= \frac{\chi(\bm{k})}{1+ \chi(\bm{k})} \frac{1}{\tNL(k)}.
\label{tsplocal}
\end{equation}
This way of writing the spectral transfer rate (in terms of the nonlinearity parameter $\chi$) makes it clear that the combination 
\begin{equation}
\sigma(\bm{k}) = \frac{\chi(\bm{k})}{1+\chi(\bm{k})}
\label{sigma}
\end{equation}
acts as a suppression factor which, when multiplying the nonlinear rate $1/\tNL$, reduces the net transfer rate due to the Alfv\'en propagation effect. We note that the suppression factor admits no singularity for any value of applied magnetic field \(B_0\), but instead \(\sigma \to 1\) when the Alfv\'en time diverges (for zero \(B_0\) or for the quasi-2D regions of the spectrum). The variation of the suppression factor in the wavector plane for our example simulation is shown in Figure \ref{fig:sigma}. 

\begin{figure}[h]
\centering
\includegraphics[scale=.56]{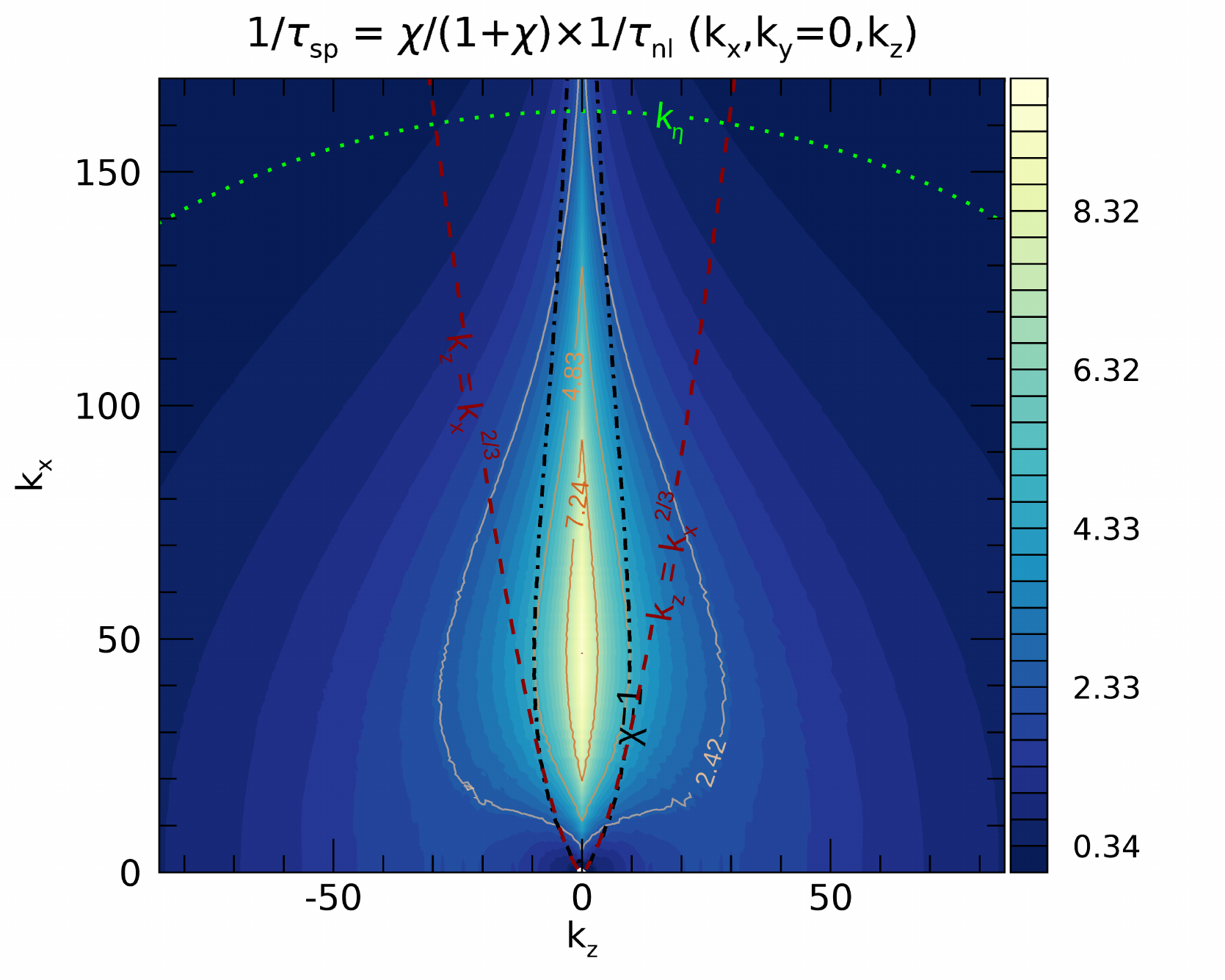}
\includegraphics[scale=.56]{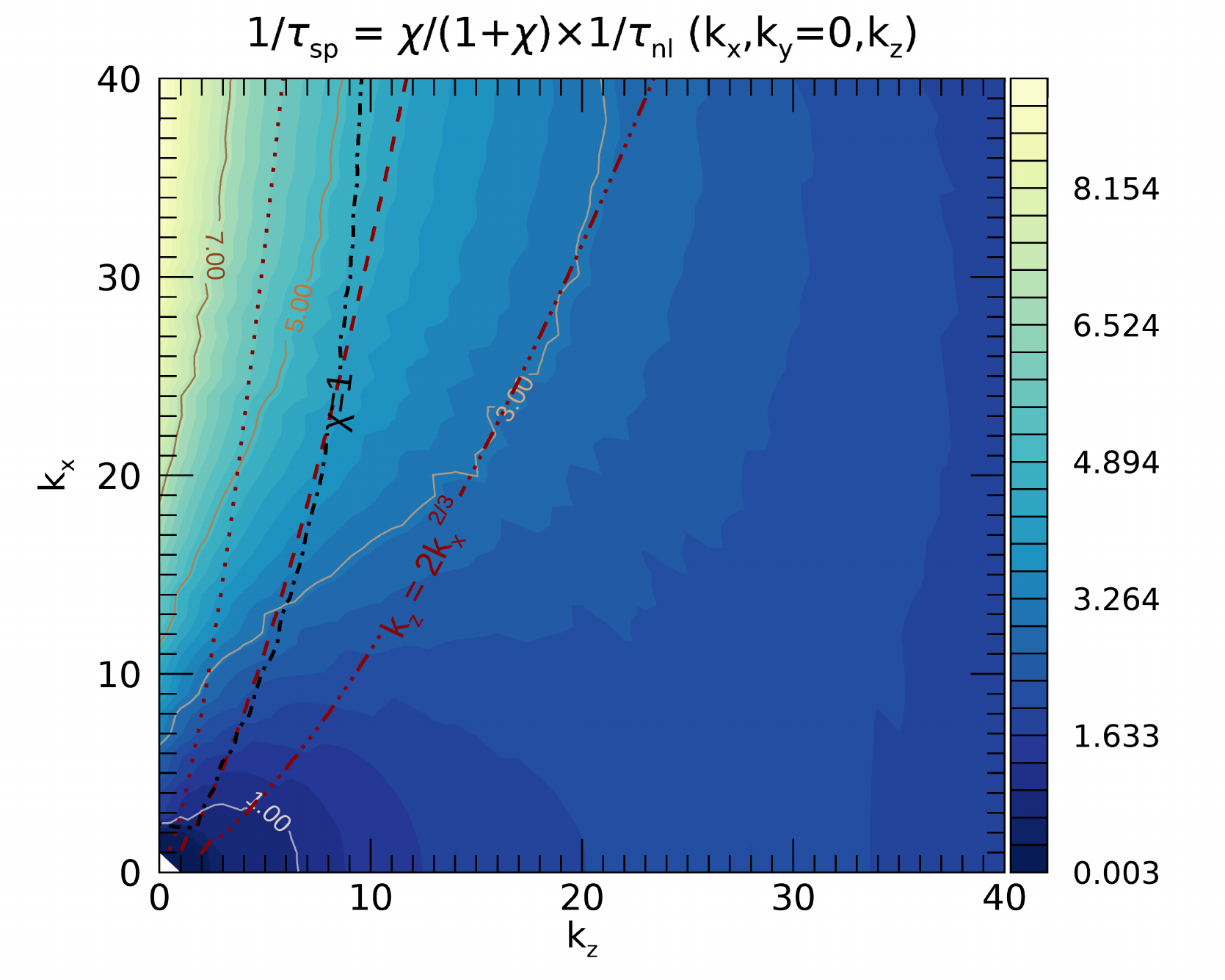}
\caption{\textit{Top}: Anisotropic spectral transfer rate \(1/\tau_\text{sp}=\tau_3(\bm{k})/\tNL^2(k) =  \frac{\chi}{\chi + 1} \frac{1}{\tNL}\), where \(\tau_\text{sp}\) is the spectral transfer time and \(\tau_3\) is the triple correlation time (see text). At \(k_z = 0\) we have \(\chi \to \infty\), and therefore we set the factor \(\frac{\chi}{\chi + 1}\) equal to one. \textit{Bottom}: A blow-up of the region near the origin. In both panels, the dissipation wavenumber and the \(\chi=1\) and Higdon curves are plotted as described in Figure \ref{fig:Emod}.}
\label{fig:inverse_t3_tnl}
\end{figure}

The above approximate formulation of the spectral transfer rate $1/\tau_\text{sp}$ achieves the desired  connection with basic timescales and permits contact with several existing theoretical frameworks; it is illustrated in Figure \ref{fig:inverse_t3_tnl}, computed from the standard simulation described above. It is evident that the very strong values of the spectral transfer rate are concentrated near the 2D plane defined by $k_z=0$, consistent with the behavior of the suppression factor in that region. In particular, this characterization holds for values of perpendicular wavenumber that were not excited initially. 

\subsection{Estimates of local strength of energy transfer}\label{sec:transfer}

The above anisotropic triple decay time is appropriate for introduction into estimation of a \emph{modal} rate of energy transfer, that is semi-local in wavevector space. This will have dimension energy per unit mass per unit time. If constructed in a physically reasonable way, this would provide a phenomenological (non-rigorous) estimate of the contributions to the total energy transfer rate due to the energy residing near a wavevector \(\bm{k}\).

There are several ways to proceed to develop  cascade rate estimates that incorporate anisotropic spectral transfer. The most formal approaches are those based on adaptations of third-order laws. \citep{politano1998GRL} An example is Ref. \citenum{stawarz2009ApJ}, which assumed a two-component slab+2D model of turbulence geometry to estimate such rates.
In principle, with sufficient data coverage, it is also possible to apply the third-order cascade law for arbitrary geometry (see e.g., Refs.  \citenum{osman2013PRL,bandyopadhyay2018epsilon})
, provided that the system is steady and homogeneous. Such approaches rely on relatively delicate third-order statistics, and do not necessarily reveal dependence on underlying physical timescales. 

In Appendix~\ref{sec:app}, beginning from first principles and adopting a minimal set of approximations, including scale locality, we develop a justification for an estimation of cascade strength in an anisotropic inertial range. Two distinct approaches are needed when there is a significant mean magnetic-field, corresponding first to the case of non-resonant transfer, and then to the case of resonant transfer. \citep{montgomery1981PoF,shebalin1983JPP} In making this distinction it becomes relevant to distinguish two separate roles of modes contributing to any particular triadic interaction: \emph{participating} modes experience an exchange of energy, while a \emph{spectator} mode acts to drive this exchange while its energy remains unchanged. \citep{oughton2006PoP,matthaeus2009PRE}

The case of non-resonant transfer, which will also be relevant when there is no substantial mean magnetic field, will be based on the timescales discussed above and a generalization of the argument leading to Equation \eqref{eq:epsilonk}. The corresponding spectral transfer estimate is 
\begin{equation}
\hat \epsilon(\bm{k})_\text{non-res} = \frac{\delta v^2(\bm{k})}{\tau_\text{sp}(\bm{k})} =
\frac{\tau_3(\bm{k})}{\tNL^2(k)} \delta v^2(\bm{k})
= \frac{\chi(\bm{k})}{1+ \chi(\bm{k})} \frac{\delta v^2(\bm{k})}{\tNL(k)}.
\label{xfernonres}
\end{equation}

For resonant transfer the argument is modified so that the nonlinear rate, and the triple correlation time of the \textit{spectator} mode is used in the approximation, rather than the corresponding timescales associated with the on-shell participating mode. \citep{matthaeus2009PRE} The spectator modes for resonant transfer are quasi-2D and have approximately zero wave frequency, so the suppression factor $\chi/(1+\chi) \to 1$ for these couplings. Furthermore, scale-local resonant transfer is not driven by the entire \(k\)-shell, but only by the 2D energy near $ \bm{B}_0 \cdot \bm{k} \approx 0$. Consequently we estimate the resonant energy transfer as 
\begin{equation}
\hat \epsilon(\bm{k})_\text{res} 
= \frac{\delta v^2(\bm{k})}{\tau_\text{nl2D}(k)}.
\label{xferres}
\end{equation}
where $\tau_\text{nl2D}(k) = [k \delta v_\text{2D}(k)]^{-1}$ is the quasi-2D on-shell nonlinear timescale. Here \(\delta v_\text{2D}(k)\) is estimated as \(\sqrt{k E_\text{2D}(k)}\) where \(E_\text{2D}\) is the contribution of 2D modes (defined by \(k_z=0\)) to the total energy in the \(k\)-shell (see Figure \ref{fig:kdiagram}). We call attention to the fact that for resonant transfer, the triple decay time is equal to the nonlinear time $\tau_\text{2D}(k)$ that drives the process.  In addition, the resonant transfer is strictly in the perpendicular direction (see Appendix~\ref{sec:app}).

\begin{figure}[h]
\centering
\includegraphics[scale=.56]{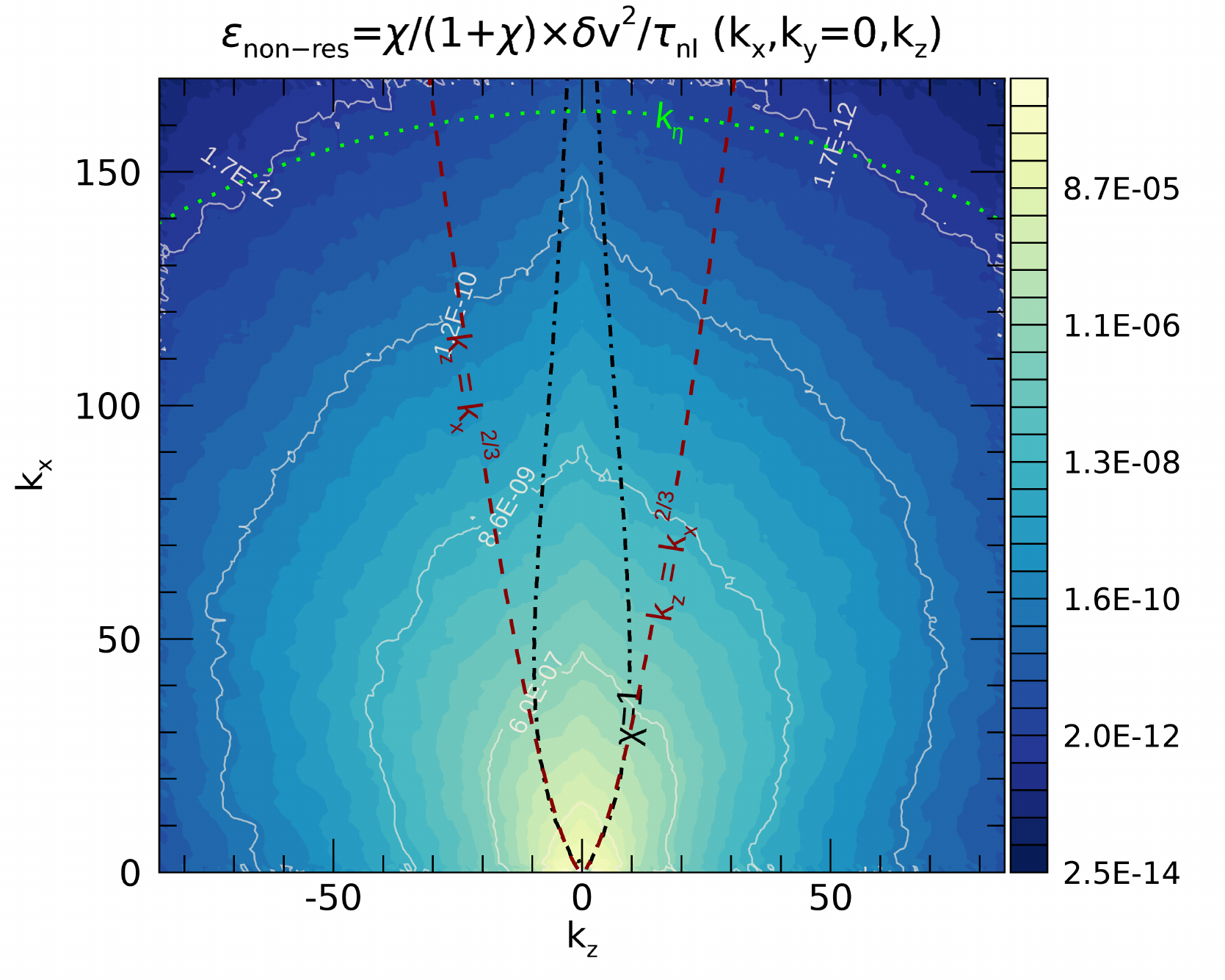}
\includegraphics[scale=.56]{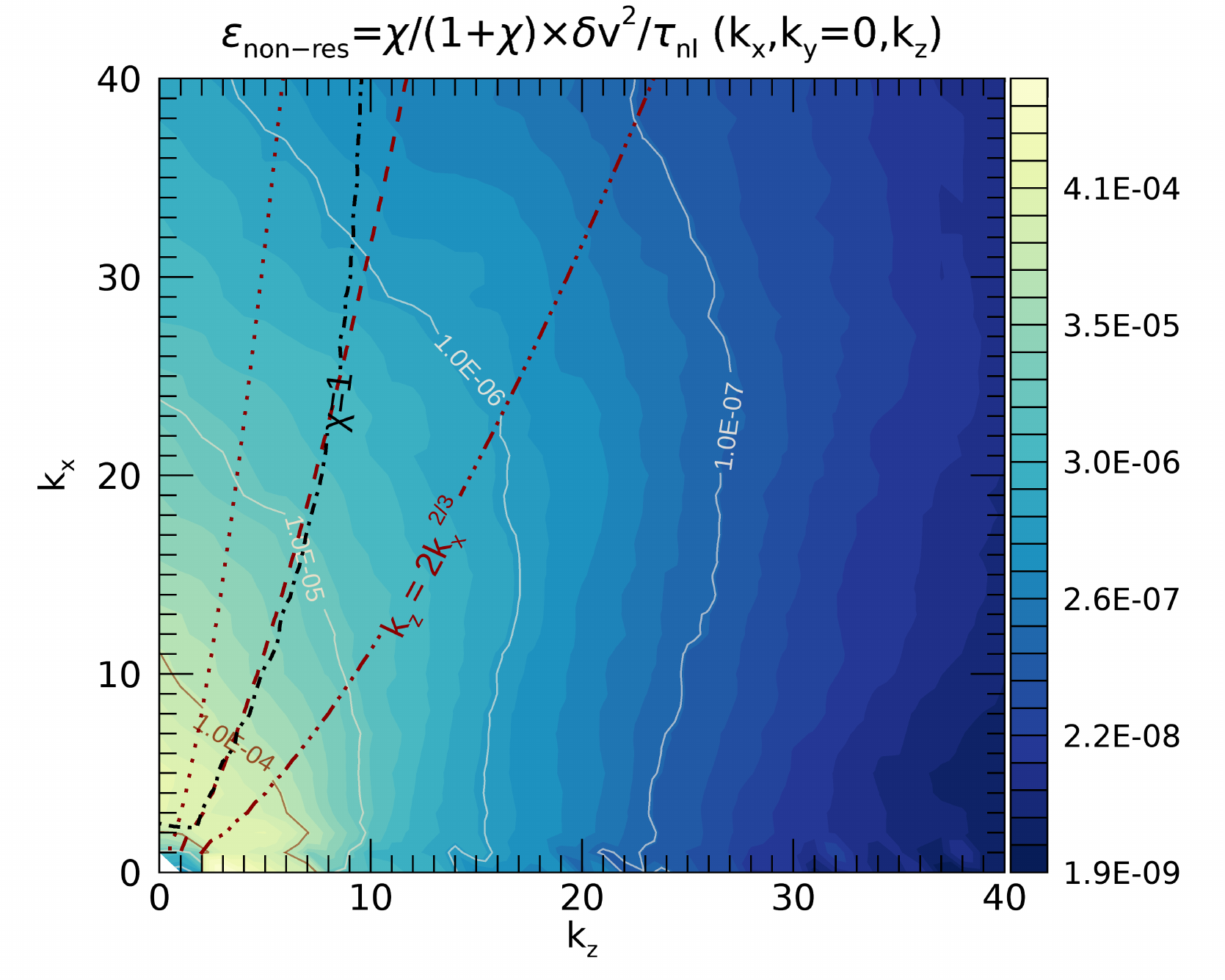}
\caption{\textit{Top}: Estimate of nonresonant energy transfer   \(\hat \epsilon(\bm{k})_\text{non-res} =\delta v^2(\bm{k})/\tau_\text{sp}^2(k) =\frac{\chi(\bm{k})}{1+\chi(\bm{k})}\frac{\delta v^2(\bm{k})}{\tNL(k)}\). At \(k_z = 0\) we have \(\chi \to \infty\), and therefore we set the factor \(\frac{\chi}{\chi + 1}\) equal to one. \textit{Bottom}: A blow-up of the region near the origin. In both panels, the dissipation wavenumber and the \(\chi=1\) and Higdon curves are plotted as described in Figure \ref{fig:Emod}.}
\label{fig:xfernonres}
\end{figure}

To illustrate these two contributions to energy transfer, we employ data from the standard simulation used above. First,  we plot estimates of the associated nonresonant contribution in Figure \ref{fig:xfernonres} in the same planar cut through the wavevector space employed above. In Figure \ref{fig:xferres} we plot the estimated resonant energy transfer using the same data  and in the same wavevector plane.

We observe in Figure \ref{fig:xfernonres} that the estimated nonresonant energy transfer, like other quantities depicted above, is anisotropic with larger values extended in the perpendicular direction. For example, the energy transfer estimate at $k_x=100$ and $k_z=0$ is larger than the value at $k_x=0$ and $k_z = 50$. Furthermore, the contours of equal estimated energy transfer differ greatly from the $\chi=1$ and 
Higdon curves that are superimposed on the figure. This is consistent with a lack of parallel transfer in regions where $\tA < \tNL$ , and stronger parallel transfer for  $\tA > \tNL$.
As a general comment, it is difficult to identify specific features of the estimated nonresonant energy transfer that correspond to either the $\chi=1$ curve or the Higdon curve.

\begin{figure}[h]
\centering
\includegraphics[scale=.5]{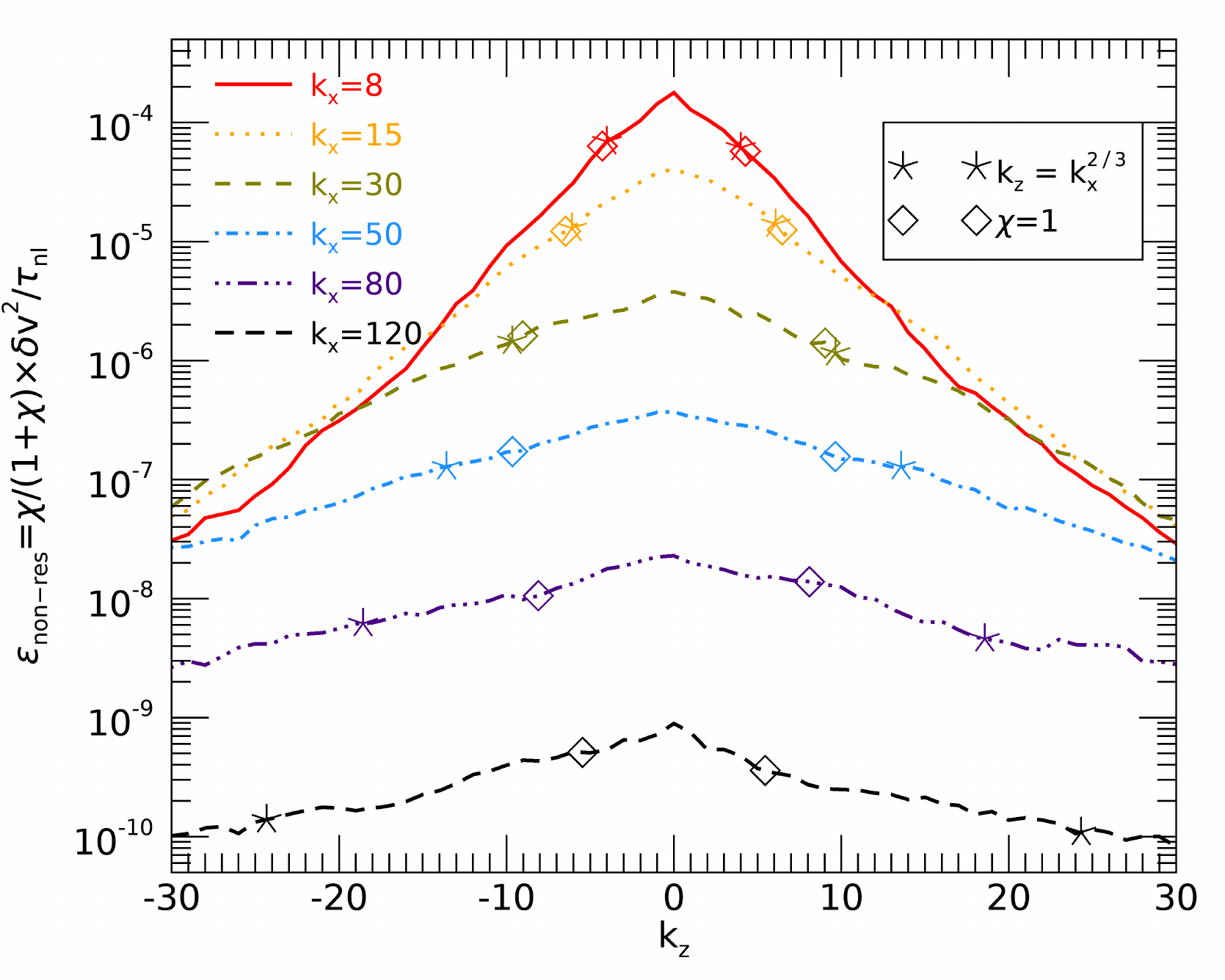}
\caption{Horizontal slices along (parallel wavenumber) \(k_z\) of the non-resonant spectral transfer rate of energy \(\hat \epsilon(\bm{k})_\text{non-res} = \chi/(\chi+1) \times \delta v^2(\bm{k})/\tNL\), through the \(k_z\dash k_x\) plane shown in Figure \ref{fig:xfernonres}, for different values of (perpendicular wavenumber) \(k_x\). The locations where the Higdon and \(\chi=1\) curves intersect these slices  are marked with `\(\ast\)' and `\(\diamond\)' symbols, respectively.}
\label{fig:xfer-slice}
\end{figure}

Figure \ref{fig:xfer-slice} shows profiles of the nonresonant energy-transfer rate estimate along lines in the $(k_x,k_z)$ plane. Each curve is for a fixed perpendicular wavenumber $k_x$ and is plotted as a function of parallel wavenumber $k_z$. The curves each show maximum values at $k_z=0$. It is therefore clear that the strongest estimated spectral transfer is found along the 2D ($k_z=0$) plane for all plotted values of perpendicular wavenumber, which are chosen to span the inertial range. In addition, each curve is annotated with a `\(\diamond\)' symbol at the (pair of) values of parallel wavenumber corresponding to the value $\chi=1$ for the nonlinearity parameter. Also indicated by `\(\ast\)' symbols are the intersections of each line with the Higdon curve (in dimensionless terms, $k_z = \pm k_x^{2/3}$). It is apparent that the values of estimated nonresonant energy-transfer at these special values of $k_z$ are systematically lower than the estimated energy transfer on the 2D axis. In fact, at $k_x=8$ and $k_x=15$ the energy transfer at the 2D plane is approximately five times greater than that found at the intersection with $\chi=1$. At higher perpendicular wavenumbers the contrast is less dramatic, but one still finds about a factor of two larger estimated energy transfer at zero parallel wavenumber as compared to the position of unit nonlinearity parameter $\chi=1$.  

The estimated resonant transfer (Figure \ref{fig:xferres}) exhibits a qualitatively similar anisotropy to that seen in the the nonresonant estimates (Figure \ref{fig:xfernonres}).  Contours of equal estimated transfer are elongated in the perpendicular direction, indicating a steeper decline in contributions to the cascade in the parallel direction. Such a distribution of turbulence activity is consistent with the real-space expectation that the cascade produces gradients that are steeper in directions perpendicular to the mean magnetic field.

%
\begin{figure}[h]
\centering
\includegraphics[scale=.56]{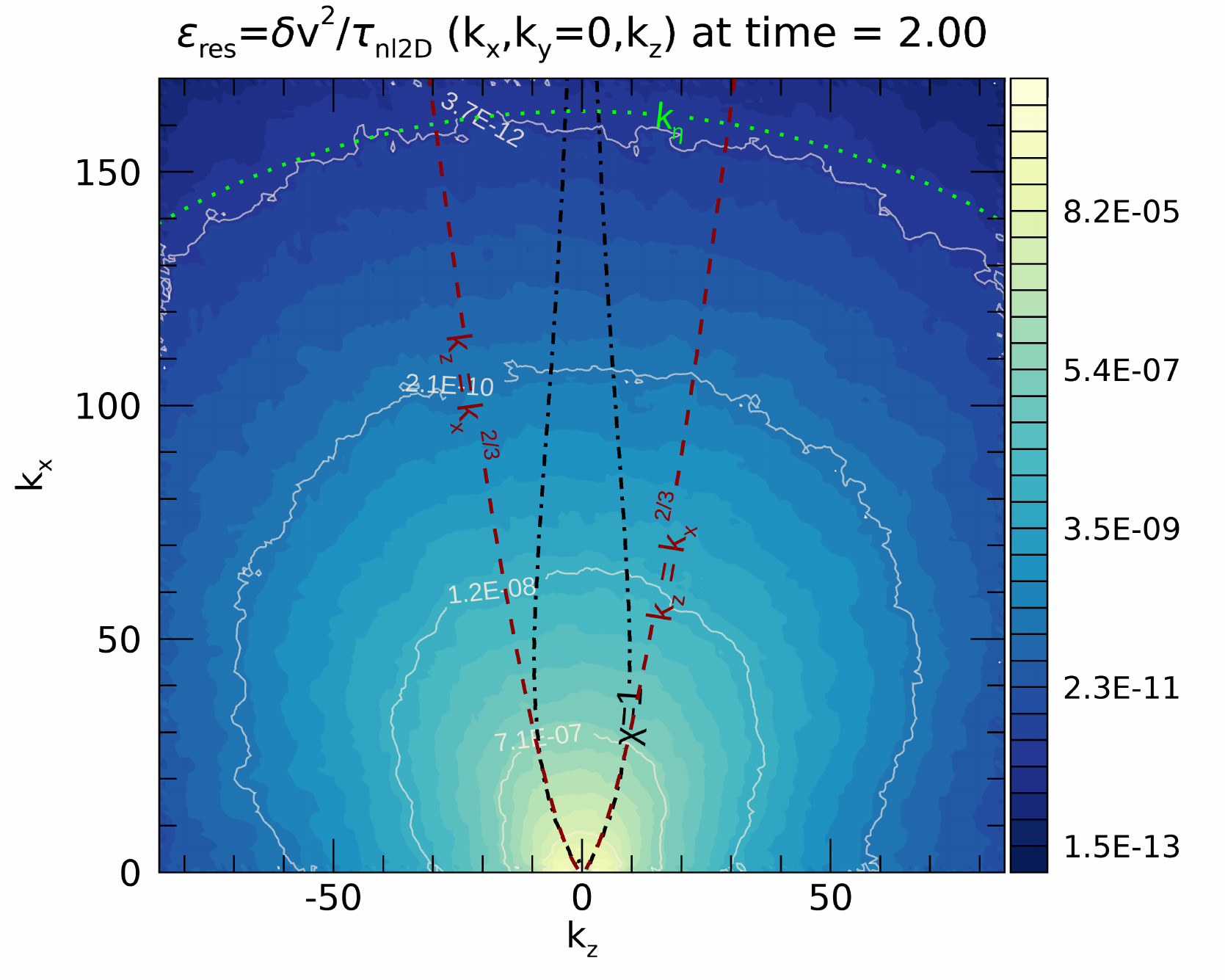}
\includegraphics[scale=.56]{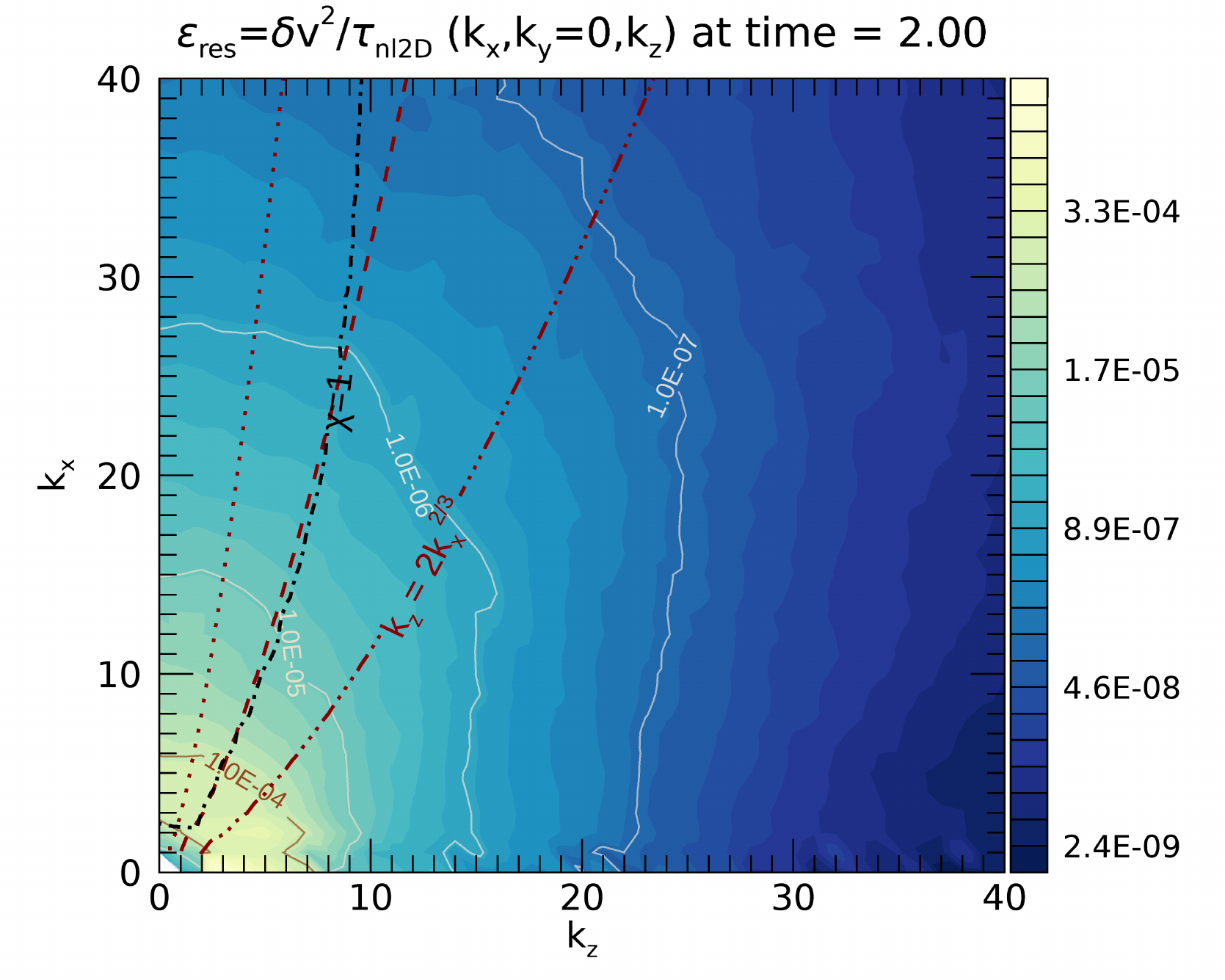}
\caption{\textit{Top}: Estimate of \textit{resonant} energy transfer 
\(\hat \epsilon(\bm{k})_\text{res} = \delta v^2(\bm{k})/\tau_\text{nl2D}(k)\), where \(\tau_\text{nl2D}(k) = 1/[k\sqrt{k E_\text{2D}(k)}]\). Only 2D modes contribute to nonlinear time on the entire $k$-shell (see text). \textit{Bottom}: A blow-up of the region near the origin. In both panels, the dissipation wavenumber and the \(\chi=1\) and Higdon curves are plotted as described in Figure \ref{fig:Emod}.}
\label{fig:xferres}
\end{figure}
\begin{figure}[h]
\centering
\includegraphics[scale=.48]{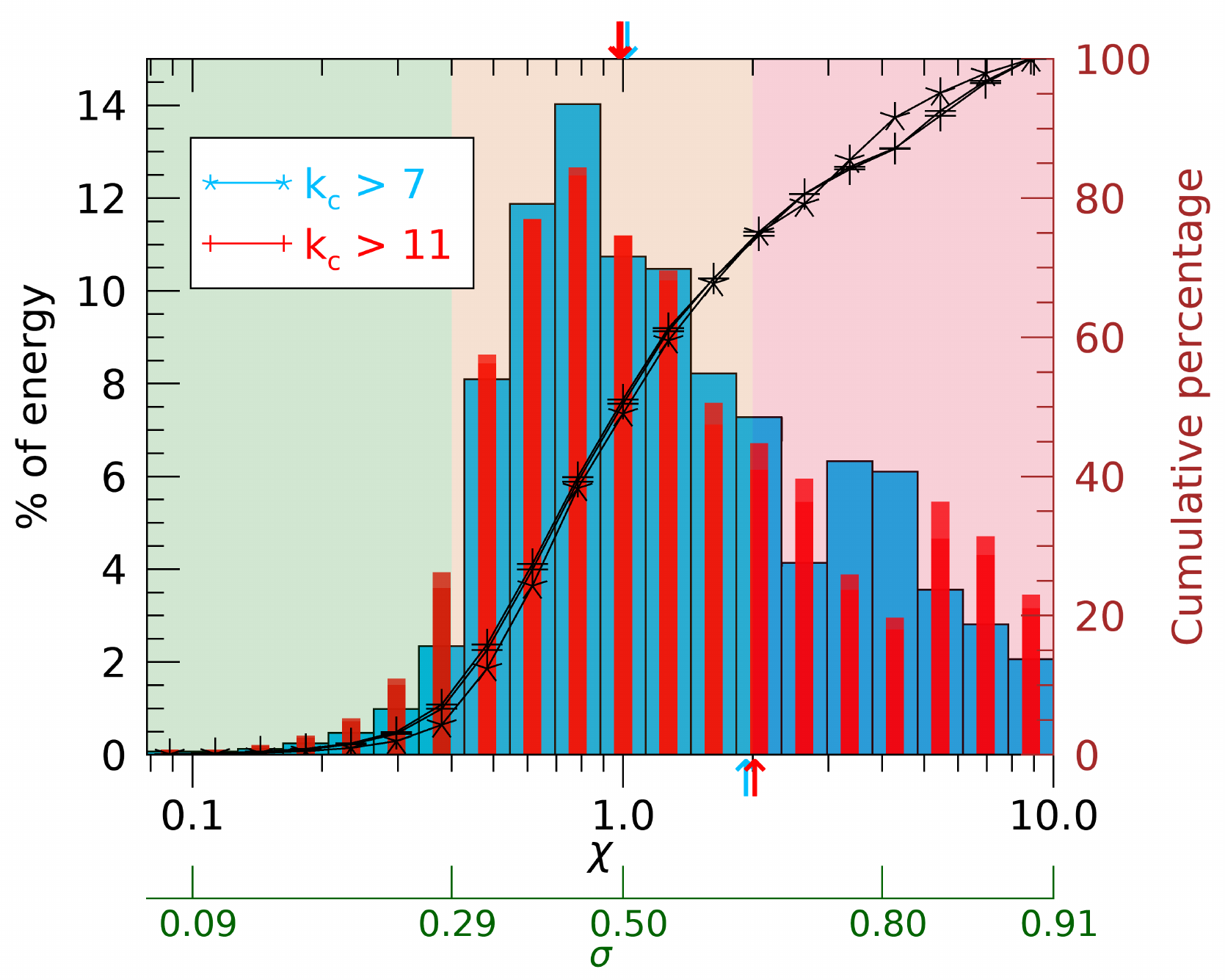}
\includegraphics[scale=.48]{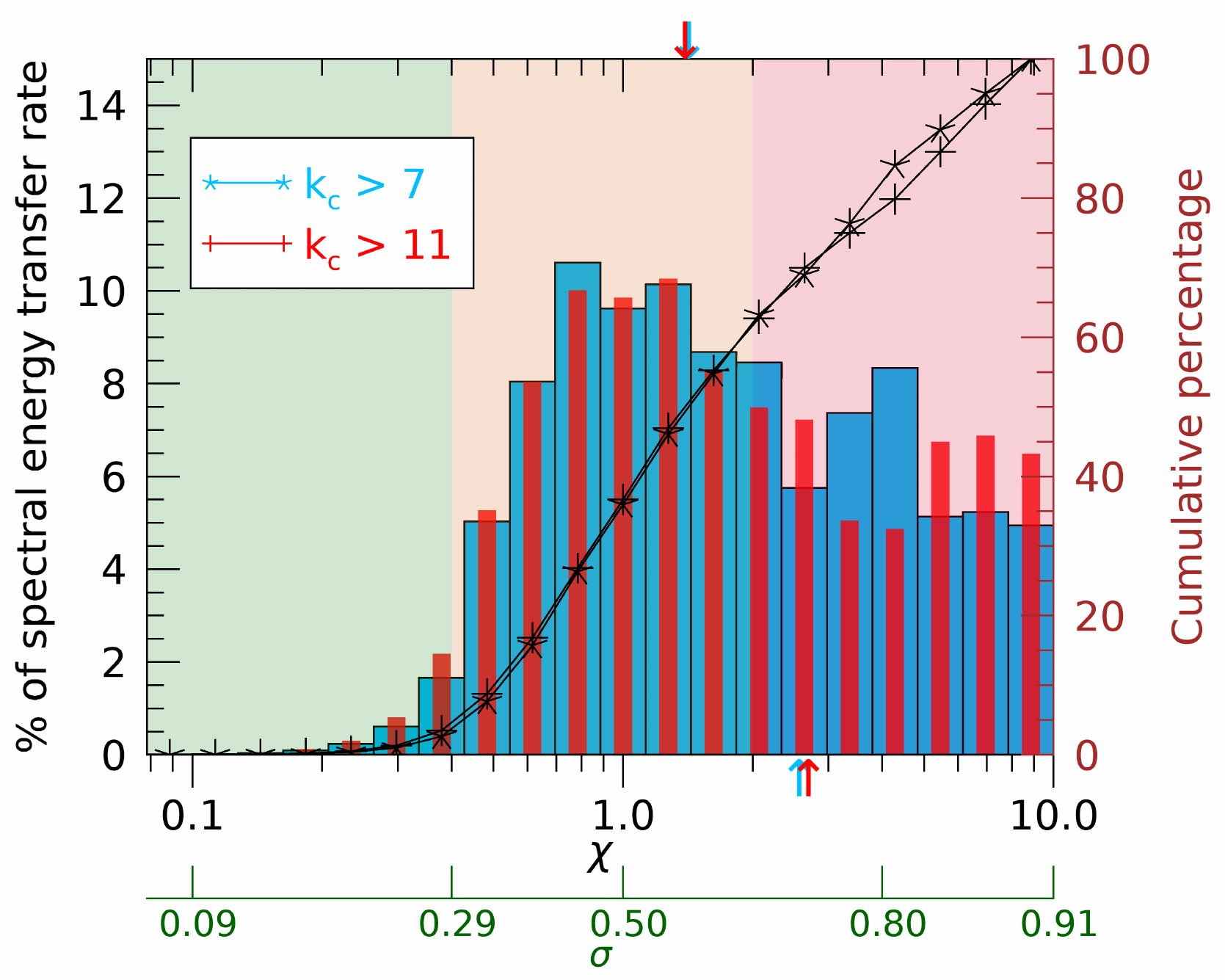}
\caption{\textit{Left:} Barplots show percentage of modal energy (left vertical axis) in bins of \(\chi\), for two values of high-pass cutoff wavenumber \(k_\text{c}\); blue bars for \(k_\text{c} > 7\) and narrow red bars for the \(k_\text{c} > 11\). The bins are equally spaced in log-scale. Curves with `\(\ast\)' and `+' symbols show cumulative percentage of energy (right vertical axis in brown color) for the two cases. Arrows (\(\uparrow\)) below the lower horizontal axis mark the weighted mean of \(\chi\) (see text). Arrows (\(\downarrow\)) above the upper horizontal axis mark the weighted median of \(\chi\) (where the cumulative percentage of energy is 50\%). Regions shaded pale green, brown, and pink represent regions of weak, moderate, and strong nonlinearity/turbulence, with the boundaries for the three regions set at \(\chi \leq 0.4,~0.4 < \chi \leq 2,\) and \(\chi > 2\), respectively. \textit{Right:} As above, with the modal energy replaced by the non-resonant spectral energy transfer rate \(\hat{\epsilon}_\text{non-res}\). Green axes at the bottom of each panel show vaues of the suppression factor \(\sigma\) corresponding to selected representative values of \(\chi\).}
\label{fig:condit}
\end{figure}

To further identify the role of the nonlinearity parameter $\chi$ in organizing the turbulence, Figure \ref{fig:condit} shows two distributions of turbulence parameters over $\chi$, including modes with \(0\le k_x\le 200\) and \(-100\le k_y,k_z\le 100\). In each case, occurrences along the 2D plane (where \(\chi\) is formally infinite) were treated as though their $\chi$ value is the same as for the nearby $k_z=1$ modes). The left panel shows the distribution of fluctuation energy in ranges (bins) of $\chi$ that have equal width in log-space. The energy in the initially populated energy-containing modes is excluded by employing high-pass wavenumber cutoffs, to avoid biasing the distribution with the (isotropic) initial data. Two values of the cutoff, $k_\text{c}>7$ and $k_\text{c}>11$, are employed, and the results for each of these is shown. It is apparent that the distribution, while peaked at a value slightly less than $\chi=1$, is also asymmetric and skewed towards larger values. As indicated on the figure, 50\% of the energy lies below (above) a value of $\chi$ that is very close to unity. However, due to the skewness of the distribution, the energy-weighted mean value of $\chi$ is approximately $2$ and changes very little for the two high-pass wavenumber thresholds. This weighted mean is computed as \(\langle\chi\rangle = \frac{\sum_i \chi_i w_i}{\sum_i w_i}\), where \(\chi_i\) is the value of \(\chi\) at the center of the \(i^\text{th}\) bin and \(w_i\) is the percentage of energy in the bin.

A similar picture emerges on inspection of the right panel of Figure \ref{fig:condit} that shows the distribution over $\chi$ of the local estimates of nonresonant energy transfer \(\hat{\epsilon}_\text{non-res}\), again for the same two high-pass wavenumber cutoffs. Once more the distribution is peaked near $\chi=1$ and here is even more strongly skewed. 50\%  of the values are found below about 1.4. The weighted mean value of $\chi$ in this case is a little greater than 2. This result is qualitatively similar to that obtained by Refs.\citenum{maron2001ApJ} and \citenum{mallet2015MNRAS}, with some possibly significant differences in the details of the simulations and in the analysis. However, the distributions shown here reveal quantitatively the energy content and spectral transfer rates prevailing in modes with different values of \(\chi\). The present diagnostics suggest that a wide range of values of nonlinearity parameter are realized in anisotropic MHD turbulence, with substantial contributions from values of \(\chi\) well above unity. These results are consistent with those from a run with isotropically polarized fluctations 
 (see Appendix~\ref{sec:app_iso}). 

As a final point we should emphasize that the treatment of spectral transfer in this section has consisted of estimating a scalar transfer rate: contributions from each point in \(k\)-space to the total energy transfer due to activity at that point. We have not made an attempt to further model this as a directional vector flux, as defined in Equation~\eqref{eq:surface} of Appendix~\ref{sec:app}. The vector flux model would be required to demonstrate steady flux across scales by integrating the normal component over a closed surface in wavevector space. 
See \cite{matthaeus2009PRE} for an anisotropic MHD model of this type based on \(k\)-space diffusion.

\section{Discussion and Conclusions}\label{sec:conclude}

In this paper we have examined in some detail the relationship between spectral anisotropy and physical timescales in a computation of MHD turbulence. This is a topic that has been extensively discussed in other studies (see the References) from a variety of perspectives and often with a motivation oriented towards establishing or promoting a particular theoretical framework. Previous works on closely related topics have often adopted approaches that we intentionally avoid here. First, in the present work we examine the simulation results in terms of simple estimates of timescales and transfer rates, reporting the analysis with a minimum of critical commentary. This stands in strong contrast to analyses that adopt approximations based on assumptions about timescales in order to derive particular models of turbulence. Examples of this would include: Montgomery and Turner's derivation of RMHD, in which dynamical timescales are assumed to be slow in comparison with Alfv\'en timescales; \citep{montgomery1981PoF} Goldreich and Sridhar's critical balance theory, which asserts that turbulence evolves such that Alfv\'en and nonlinear timescales remain nearly equal; \citep{goldreich1995ApJ} and \emph{weak turbulence} theory, which requires that Alfv\'en times be shorter than nonlinear times.\citep{galtier2000JPP} We make none of these approximations here. Second, as noted in earlier sections, we have not adopted the most rigorous available frameworks for quantifying spectral transfer. For the problem at hand this would correspond to the several possible forms of the Kolmogorov--Yaglom--Politano--Pouquet \emph{third-order} laws that are available for application to MHD simulation.\citep{politano1998GRL,
osman2011PRL_third,verdini2015ApJ} These are extremely powerful tools but they do not provide direct information about the available physically-relevant timescales. Such information is contained in higher-order correlations. Finally, we remind the reader that the present results are based on analysis of a single snapshot from a simulation initialized with toroidally polarized fluctuations, and we have made no attempt to claim any form of universality concerning the results. 

To confirm the robustness of our results we repeated our analyses for a simulation initialized with isotropic fluctuations (with toriodal + poloidal polarization), and obtained similar results (see Appendix~\ref{sec:app_iso}). Therefore it is our impression that the results presented here are of a fairly typical character for the class of initial data that we have examined. This corresponds, roughly speaking, to low cross-helicity, incompressible ``Alfv\'enic'' turbulence, and, within limitations of incompressible MHD, does not differ greatly from parameters often chosen to simulate solar-wind-like conditions at 1 AU (e.g., Refs. \citenum{goldstein1995araa,tu1995SSRv,
bruno2013LRSP}).

The main results of the paper are the estimates of the anisotropic distributions of estimated spectral transfer rates and rates of energy transfer (shown in Figures \ref{fig:inverse_t3_tnl}--\ref{fig:xferres}), as well as the distributions of energy and its transfer rate across the  nonlinearity parameter $\chi$ and the suppression factor $\sigma$ (shown in Figure \ref{fig:condit}).  As mentioned above, these results pertain only to the specific case of homogeneous MHD turbulence with a mean magnetic field of moderate strength. The results may be summarized succinctly with the statement that the distributions of estimated transfer rates are relatively featureless. There are no strong enhancements in any part of \(k\)-space beyond a local peak of nonlinear activity near the 2D axis, which is globally perpendicular to the direction of the externally applied field. This is broadly consistent with expectations based on derivations of the RMHD model, \citep{montgomery1982PhysScrip,zank1992JPP,oughton2017ApJ} and, at least superficially, appears to stand in contrast to certain interpretations of critical balance that anticipate a special role for modes having equal nonlinear and Alfv\'en timescales. Further examination of the relationship of nonlinear and Alfv\'en timescales was performed by looking at the distributions of energy and spectral energy transfer in bins of   nonlinearity parameter $\chi=\tA/\tNL$ and the suppression factor $\sigma=\chi/(1 + \chi)$. The distributions are peaked near \(\chi=1\) (or equivalently, \(\sigma=0.5\)), but are skewed towards larger values of \(\chi\) corresponding to strong nonlinearity. The broad distributions of nonlinearity parameter admit numerous values and an average (weighted by energy and its spectral transfer rate) that exceeds unity. It is therefore difficult to argue in favor of theoretical developments, such as certain interpretations of critical balance, that postulate that $\chi = 1$ is a limiting maximum value (e.g., Refs. \citenum{goldreich1995ApJ,goldreich1997ApJ,
maron2001ApJ,
schekochihin2009ApJS182,tenbarge2012PoP,
mallet2015MNRAS}).

A brief comment on local (as opposed to global) anisotropy is in order. It is well known \citep{cho2000ApJ,Milano2001PoP,Matthaeus2012ApJ} that analysis of anisotropy using structure functions parallel and perpendicular to locally-computed average magnetic fields gives results with higher anisotropy. It is also established that this effect is sensitive to phase coherence or intermittency, and disappears when the fields are ``gaussianized''. This enhancement is therefore fundamentally of  higher order than classical second-order spectra (see Ref. \citenum{Matthaeus2012ApJ}). However, our interest in the present paper is in classical spectra (and underlying correlation functions) and these are only well defined in a fixed coordinate system in which the preferred reference direction is fixed.

The interested reader might find additional useful information about effects that are studied using the local, random coordinate system in, e.g., Refs. \citenum{maron2001ApJ,
schekochihin2009ApJS182,mallet2015MNRAS}. The issue of the appropriate mean fields to use when analyzing spectral anisotropy has received substantial attention (e.g., Refs. \citenum{cho2000ApJ,horbury2008PRL,
schekochihin2009ApJS182,Podesta2009ApJ,
isaacs2015JGR120,Wang2016ApJ,
Gerick2017ApJ,telloni2019ApJ,Wu2020ApJ}). Clearly, there are both technical and conceptual aspects to address regarding this issue and we defer fuller consideration to future work.

While we have attempted to avoid adopting bias towards or against any specific theoretical approach, it seems clear that the present results favor a somewhat simplified description of anisotropy in MHD associated  with a mean magnetic field. The idea inherent in the derivation of RMHD, \citep{zank1992JPP} namely that the 2D or quasi-2D modes represent the core nonlinearities of a turbulent system, appears to be largely consistent with this single detailed numerical experiment. Both resonant and nonresonant estimated transfer rates point toward maintenance of anisotropic spectra with strong perpendicular real-space gradients relative to the (suppressed) parallel gradients. The critical-balance condition, $\chi= 1$, appears to be better interpreted as an order-of-magnitude estimate of the extent of the spectra in the parallel direction, at least in this case where the measurements are made relative to a global magnetic field. In this way most of the realizable implications of critical balance point to a dynamical description in terms of  quasi-2D or Reduced MHD. Such a description of anisotropic MHD turbulence is a simplification relative to the chain of reasoning leading to critical balance models, and we await further analyses that provide support for the present viewpoint, or alternatives. A more complete discussion of the role of critical balance ideas in MHD turbulence is under review (Oughton \& Matthaeus 2020). Related future work could examine similar distributions of the timescales in high cross helicity simulations as well as in spacecraft observations.\citep{matthaeus2014apj,telloni2019ApJ} 

\begin{acknowledgments}
We thank R. Bandyopadhyay for useful discussions. This research was supported in part by NASA Heliospheric Supporting Research grants NNX17AB79G, 80NSSC18K1210, and 80NSSC18K1648. The data that support the findings of this study are available from the corresponding author upon reasonable request. 
\end{acknowledgments}

%
\appendix
\section{Spectral Transfer and Scale Locality}\label{sec:app}
Here we develop some background to motivate our approach to estimating semi-local contributions in wavevector space to spectral transfer in anisotropic MHD turbulence. The spectral density of energy \(S(\bm{k})\) evolves in incompressible MHD according to an equation of the same formal structure as the analogous equation for the spectral density in incompressible hydrodynamics.  Note that \(S(\bm{k})\) is defined as the trace of the energy spectrum tensor, which in turn is defined in terms of the Fourier transform of the two-point correlation function of the turbulent fields (see, e.g., Refs. \citenum{Batchelor1953book,matthaeus1982JGR}). Specifically, the energy spectrum evolves as (e.g., Chapter 6 of Ref. \citenum{lesieur2008book})
\begin{equation}
\frac{\partial S(\bm{k})}{\partial t} 
= N(\bm{k}) - D(\bm{k}),
\label{eq:dSdt}
\end{equation}
where \(N(\bm{k}) = \int \d\bm{p} \, \d\bm{q} \, T(\bm{k},\bm{p},\bm{q})\) is the nonlinear term representing the net effect of all triadic interactions on the energy density at wavevector \(\bm{k}\), \(T\) is the Fourier-space triple correlation, and 
$D$ is the dissipation term. Each term in Equation~\eqref{eq:dSdt} is time dependent.

Energy conservation in the absence of dissipation corresponds to the property that
\begin{equation}
\int \d\bm{k} \, N(\bm{k}) = \int \d\bm{k} \, \d\bm{p} \, \d\bm{q} \, T(\bm{k}, \bm{p},\bm{q}) = 0.
\label{eq:NT}
\end{equation}
We have neglected possible sources at very low $|\bm{k}|\sim 1/L$, while in the usual way the dissipation is assumed to be effective only for very large $kL$. Then there exists an inertial range in which energy is conserved by nonlinearities, even as it is transferred from scale to scale. The net dissipation is $\epsilon = \int \d\bm{k}\, D(\bm{k})$ and in steady state this is equal to the transfer rate across any sphere of radius $k$ lying in the inertial range. 

We gain insight by integrating the equation for the spectral density Equation \eqref{eq:dSdt} over all $|\bm{k}| < k^*$, and 
defining 
\begin{equation}
E_<(k^*) = \int_0^{k^*} \d{q} \int q^2 \d\Omega_q \,S(\bm{q})
\\
\textrm{  \,\,\,   and  \,\,\,     }
E_>(k^*) = \int_{k^*}^\infty \d{q} \int q^2 \d\Omega_q \, S(\bm{q}),
\label{eq:Epm}
\end{equation}
where the integral is expressed in spherical polar coordinates and \(\d\Omega=\sin\theta \, \d\theta \d\phi\) is the differential solid angle. Ignoring dissipation, it is clear that 
\begin{equation}
\frac{\d E_<}{\d t} + \frac{\d E_>}{\d t} = 0.
\label{eq:pmconserve}
\end{equation}
In solving Equation \eqref{eq:dSdt} we are confronted with the classical closure problem of turbulence. The second-order quantity $S$ depends on the third-order correlation $T$. Further development would show that the time evolution of $T$ depends on fourth-order correlations, and so on. To proceed, various approximations can be made to solve for, or bring closure to, this hierarchy (e.g., Ref.\citenum{edwards1964JFM}).

The principal complication lies in the nonlinear term described in Equation \eqref{eq:NT}, which, through the definition of the triple correlation $T$, depends on convolutions of the (schematic) form
\begin{equation}
N(\bm{k}) \sim \int \d\bm{p} \, \d\bm{q} \, \delta(\bm{p+q-k})
v(\bm{k}) v(\bm{p}) v(\bm{q}).
\label{eq:conv}
\end{equation}
Here \(\delta (\bm{p+q-k})\) is a Dirac delta function, and so only triads with \(\bm{p+q} = \bm{k}\) contribute to the integral. The associated \textit{triadic} interactions have been studied and classified in a number of studies. \citep{orszag1973lectures,kraichnan1980RPPh,domaradzki1990PoF,
waleffe1992PoF,verma2005PoP,matthaeus2009PRE,teaca2011PoP}

The development of models for the third-order correlations $T$ is a principal goal of turbulence theory. Rigorous treatments are difficult, and usually remain inexact (e.g., Refs. \citenum{orszag1973lectures,mccomb1990physics}). A useful approach, adopted here, is to justify and adopt phenomenological models for the nonlinear effects, which includes the physics of the cascade, and appropriate approximations.  

\begin{figure}
\centering
\includegraphics[scale=.75]{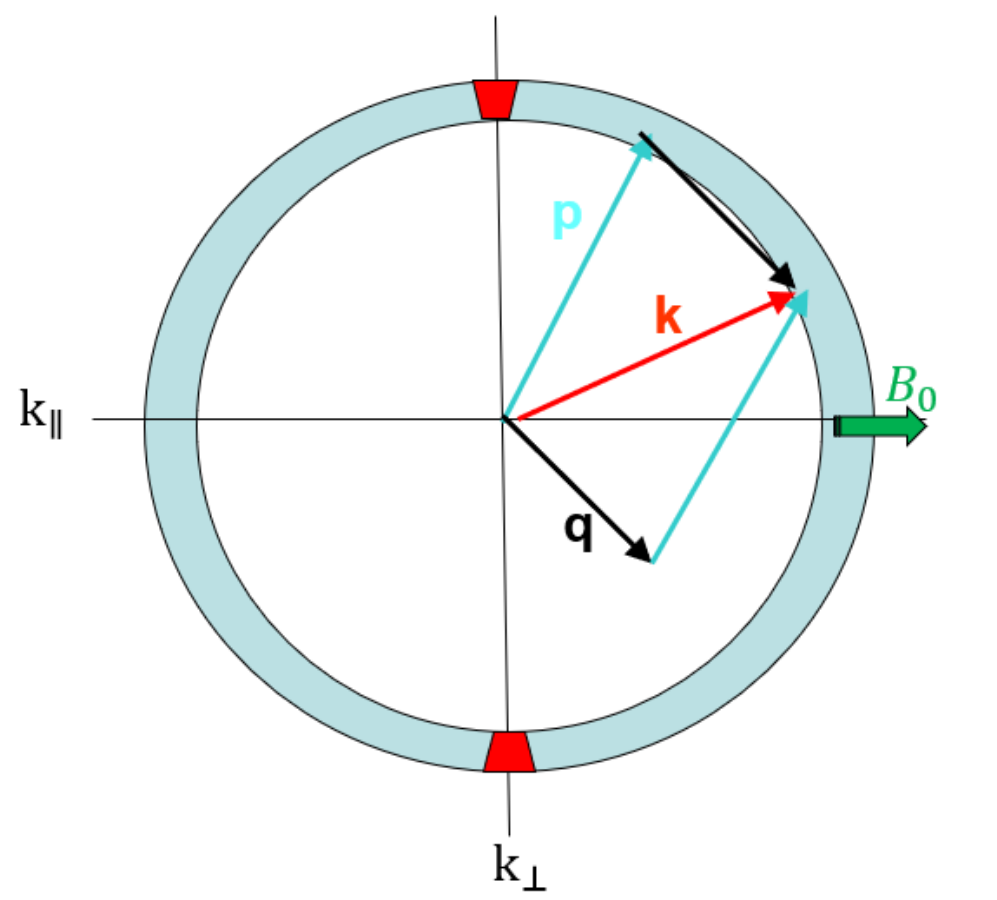}
\caption{Diagram of a 2D cut through \(k\)-space,  indicating locality of triadic interactions on (near) a shell of radius \(k\). Wavevectors for the quasi-2D modes \citep{oughton2005NPG} lying on this shell terminate in the region shaded in red.}
\label{fig:kdiagram}
\end{figure}

To guide our reasoning, we refer to Figure \ref{fig:kdiagram}, which schematically shows a 2D cut (in the \(k_\parallel\)-\(k_\perp\) plane) through 3D wavevector space. A set of wavevectors corresponding to a triadic interaction in the inertial range is illustrated. In each triadic interaction in incompressible MHD, there are spectator modes, and exchange modes. The spectator modes (say, \(\bm{q}\) in Figure \ref{fig:kdiagram}) induce energy transfer between the exchange modes, while the spectator mode energy remains unchanged. The spectator wavevector therefore indicates the (unsigned) direction in which a particular triad induces transfer. This property is general \citep[see][]{oughton2006PoP,matthaeus2009PRE} and is useful for classifying different types of triads. Note that because the net spectral transfer into \(S(\bm{k})\) is an integral over two sets of wavevectors, \(\bm{q}\) and \(\bm{p}\), the triad depicted in Figure \ref{fig:kdiagram} actually corresponds to two triads: one where \(\bm{q}\) is the spectator mode and another where \(\bm{p}\) is the spectator mode.

The idea of \textit{locality in scale} is a powerful assumption in Kolmogorov theory that is supported, even  if inexactly, in numerical experiment and theory. \citep{mccomb1990physics,domaradzki1990PoF,verma2005PoP,
alexakis2007PRE,aluie2010PRL} The basic idea is that the net transfer from wavenumbers $<k$ to wavenumbers $>k$ is dominated by triadic interactions involving wavevectors that do not differ (in magnitude) greatly from $k$ itself. This can be visualized in Figure \ref{fig:kdiagram} as consisting of cases in which the interactions occur among \(\bm{k},~ \bm{p},\) and \(\bm{q}\), all having approximately the same magnitude, \(k \sim p\sim q\). That is, all three participating wavevectors, or perhaps two of these (``Modified locality''; see Ref. \citenum{matthaeus2009PRE}), lie on or near the shell with radius $k$.

The locality property stands in contrast to the general case of triadic interactions (Equations \eqref{eq:NT} and \eqref{eq:conv}) that may depend on spectral amplitudes that are greatly distant both from the direction of \(\bm{k}\) or from the shell with \(|\bm{k}|=k\). Those that are distant are nonlocal interactions. Ignoring the nonlocal effects and adopting the approximation of scale locality, one may interpret the inertial range energy-balance Equation \eqref{eq:pmconserve} as the conservative exchange of energy across the boundary at wavenumber $k$. In that case, inertial range conservation may be expressed as 
\begin{equation}
N(\bm{k}) = \nabla_{\bm{k}} \cdot \bm{F}(\bm{k}).
\label{eq:div}
\end{equation}
With this approximation and definition, both locality and conservation are guaranteed in the inertial range. Note that the \textit{vector} flux \(\bm{F}\) admits functional dependence on correlations at wavevectors other than \(\bm{k}\), but by locality the dependence is dominated by wavevectors of similar magnitude. Then, in terms of the flux, the time rate of change of energy within a sphere of radius $k$ is 
\begin{equation}
\frac{\d E_<}{\d t}(k) = \int_{q<k} \d\bm{q} \, \nabla_{\bm{q}} \cdot \bm{F}(\bm{q})
=  k^2 \int d\Omega_{\bm{k}} \, F_k(\theta_k,\phi_k)
\label{eq:surface}
\end{equation}
where the spherical polar representations of the wavevectors \(\bm{k} = (k,\theta_k,\phi_k)\) and \(\bm{q} = (q,\theta_q,\phi_q)\) are employed.  

The approximations leading to the expression Equation \eqref{eq:surface} did not include the assumption of isotropy. Nevertheless we see that the rate of transfer of energy to all wavevectors \(\bm{q}\) having \(|\bm{q}| > k\) is determined by the conditions on the shell \(|\bm{q}| = k\). This is crucial for motivating the phenomenological estimates of spectral transfer in the main text. In particular, a model of local spectral transfer will specify the surface flux \(F_k =  \frac{\bm{k}}{k} \cdot \bm{F}_{\bm{k}}\) and requires consideration only of the physical parameters in the shell with radius \(k\). However, it is clear that physical insight is required to understand how $F_k$ will 
depend on contributions from different parts of the shell.  

In this regard it is also possible to examine types of triads, allowing several classes of interaction to be identified. \citep{oughton2006PoP,matthaeus2009PRE} Additional considerations include the vector polarization of the fluctuations, and the comparison of the scale-local turbulence amplitude to the mean field strength. An important factor that enters the modeling of triadic interactions is whether interactions are \textit{resonant interactions} or nonresonant interactions, when turbulence occurs in the presence of a sufficiently strong externally supported uniform or very large scale magnetic field. Most interacting triads are affected by this (effectively) DC magnetic field, in that the strength of the transfer they induce is suppressed, as described by the factor $\sigma=\chi/(1+\chi)$ in Equation~\eqref{xfernonres}. However, there is a particular class of triads that are unaffected by the large-scale field and are resonant interactions; that is the set of triads that include one (or all three) participating wavevector that lies in the plane perpendicular to the field \(\bm{B}_0\). Such modes have ``zero frequency'' and the triads that involve these ``2D'' modes exchange energy but they do so without changing the associated Alfv\'enic frequency of the affected fluctuations. These correspond to resonant interactions in an iterative weak turbulence approach. \citep{shebalin1983JPP,grappin1986PoF,oughton1994JFM} In designing a phenomenological description of scale-local transfer it is necessary to properly distinguish the frequency changing non-resonant interactions from the frequency-preserving resonant interactions. \citep{oughton2006PoP}

\section{Results from an Initially Isotropic Run}\label{sec:app_iso}

\begin{figure}[h]
\centering
\includegraphics[scale=.48]{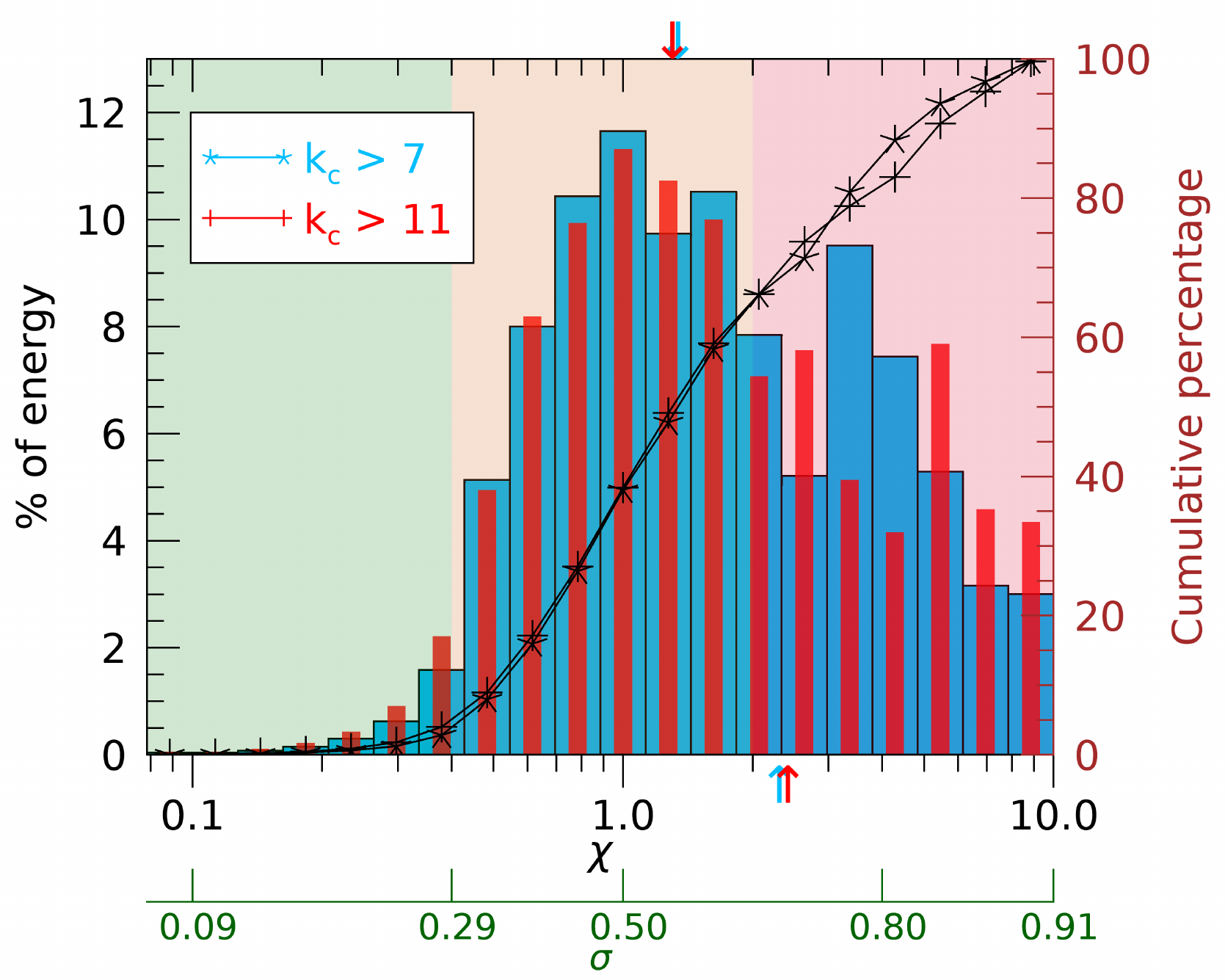}
\includegraphics[scale=.48]{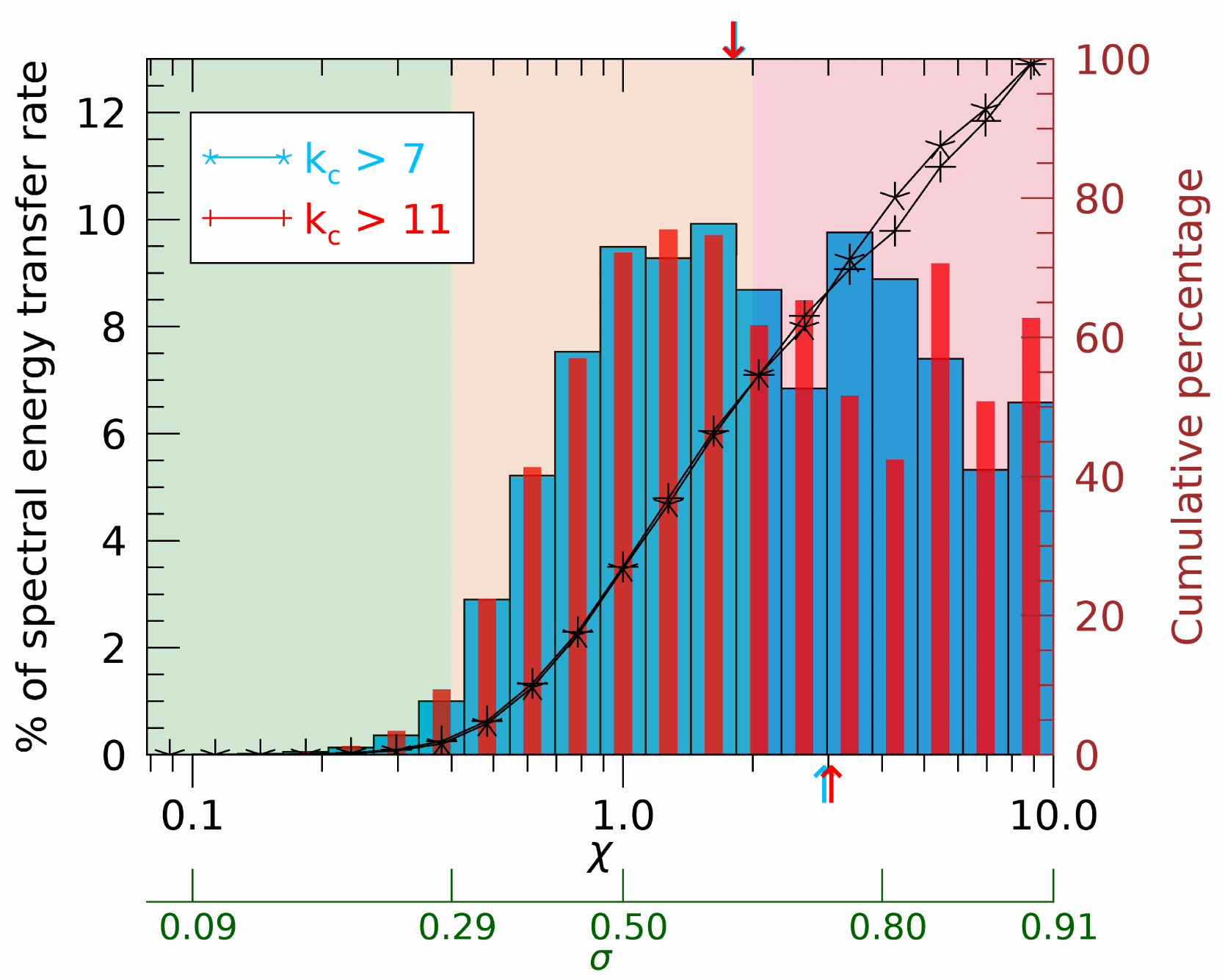}
\caption{\textit{Left:} Barplots show percentage of modal energy in bins of \(\chi\). \textit{Right:} The modal energy replaced by the spectral energy transfer rate \(\sigma  E_\text{modal}(\bm{k})/\tNL(k)\). These results are from a run with initially isotropic fluctuations (see text). All other details follow Figure \ref{fig:condit}.}
\label{fig:condit_iso}
\end{figure}

In order to examine whether the results discussed above are biased by our choice of initial polarization we carried out an additional run. The results shown in the main body of the paper are obtained for initially toroidal (transverse) polarizations (the results leading to Figure~\ref{fig:condit}), while the results shown in Figure~\ref{fig:condit_iso} were obtained for initially isotropic polarization, while all other simulation details are identical. One can observe that the results are very similar, although the energy and spectral-transfer in quasi-2D regions appears to be slightly more significant in the isotropic case. Since there are only minor differences, it is reasonable to conclude that modifying the initial polarization does not produce a large effect, and in particular, definition of the nonlinear time to take polarization anisotropy into account would make little difference.

%




%


\end{document}